\documentclass[fleqn,usenatbib]{mnras}

\usepackage{newtxtext,newtxmath}

\usepackage{booktabs}
\usepackage{multirow}
\usepackage{orcidlink}

\usepackage[T1]{fontenc}

\DeclareRobustCommand{\VAN}[3]{#2}
\let\VANthebibliography\thebibliography
\def\thebibliography{\DeclareRobustCommand{\VAN}[3]{##3}\VANthebibliography}
\usepackage{ragged2e}

\usepackage{graphicx}	
\usepackage{amsmath}	
\usepackage{amssymb}	

\title[New X-ray ALP constraints with H1821+643]{\centering{New constraints on light Axion-Like Particles using \textit{Chandra} Transmission Grating Spectroscopy of the powerful cluster-hosted quasar H1821+643}}

\author[J. Sisk Reynés et al.]{Júlia Sisk Reynés \orcidlink{0000-0003-3814-6796},$^{1}$\thanks{Contact details: jms332@cam.ac.uk.}
James H. Matthews \orcidlink{0000-0002-3493-7737},$^{1}$
Christopher S. Reynolds \orcidlink{0000-0002-1510-4860},$^{1}$
Helen R. Russell \orcidlink{0000-0001-5208-649X},$^{2}$
\newauthor
Robyn N. Smith \orcidlink{0000-0001-5626-5209},$^{3}$
and M. C. David Marsh \orcidlink{0000-0001-7271-4115}$^{4}$
\\
$^{1}$Institute of Astronomy, University of Cambridge, Madingley Road, Cambridge CB3 OHA, UK\\
$^{2}$School of Physics and Astronomy, University of Nottingham, Nottingham NG7 2RD, UK \\
$^{3}$ Dept. of Astronomy, University of Maryland, College Park, MD 20742, USA\\
$^{4}$ The Oskar Klein Centre, Department of Physics, Stockholm University, Stockholm 106 91, Sweden\\ 
}

\date{\today}

\pubyear{2021}

\begin{document}
\label{firstpage}
\pagerange{\pageref{firstpage}--\pageref{lastpage}}
\maketitle

\begin{abstract}
Axion-Like Particles (ALPs) are predicted by several Beyond the Standard Model theories, in particular, string theory. In the presence of an external magnetic field perpendicular to the direction of propagation, ALPs can couple to photons. Therefore, if an X-ray source is viewed through a magnetised plasma, such as a luminous quasar in a galaxy cluster, we may expect spectral distortions that are well described by photon-ALP oscillations. We present a $571 \ \mathrm{ks}$ combined High and Low Energy Transmission Grating (HETG/LETG) \textit{Chandra} observation of the powerful radio-quiet quasar H1821+643, hosted by a cool-core cluster at redshift $0.3$. The spectrum is well described by a double power-law continuum and broad$+$narrow iron line emission typical of type-1 Active Galactic Nuclei (AGN), with remaining spectral features $< 2.5\%$. Using a cell-based approach to describe the turbulent cluster magnetic field, we compare our spectrum with photon-ALP mixing curves for 500 field realisations, assuming that the thermal-to-magnetic pressure ratio $\beta$ remains constant up to the virial radius. At $99.7\%$ credibility and taking $\beta=100$, we exclude all couplings $g_\mathrm{a\gamma} > 6.3 \times 10^{-13} \ {\mathrm{GeV}}^{-1}$ for most ALP masses $< 10^{-12} \ \mathrm{eV}$. Our results are moderately more sensitive to constraining ALPs than the best previous result from \textit{Chandra} observations of the Perseus cluster, albeit with a less constrained field model. We reflect on the promising future of ALP studies with bright AGN embedded in rich clusters, especially with the upcoming \textit{Athena} mission. 
\end{abstract}


\section{Introduction}
\label{sec:Introduction}
\justifying
The Standard Model (SM) of Particle Physics has been extremely successful at describing a large range of physical phenomena below the Fermi or electroweak scale, $\sim 240 \ \mathrm{GeV}$. However, the apparent conservation of Charge-Parity (CP) symmetry in the strong sector, first suggested by the small upper bound on the neutron electric dipole moment, $\lesssim 10^{-21} \ e\cdot \mathrm{cm}$ \citep[][]{PhysRev.184.1660_ElectricDipoleNeutronMoment_Predictions1969}, is one of the several findings that the SM alone fails to describe.

Peccei-Quinn (PQ) symmetry was proposed as a Beyond the SM (BSM) extension to prevent CP violation by the strong force \citep[][]{PecceiQuinnOriginal}. The spontaneous breaking of PQ symmetry results in a Goldstone boson \citep[][]{FWilczek_QCDaxion, SWeinberg_QCDaxion}.

The QCD axion, a pseudo-scalar, long-lived Goldstone boson emerging from PQ symmetry is the leading solution to the strong CP-problem, since it replaces the CP violating term in the global Lagrangian by a dynamical CP conserving axion field. The properties of the QCD axion are set by the scale at which PQ symmetry is spontaneously broken, which yields a proportionality relation between the axion mass and its coupling to electromagnetism. However, QCD axions \citep[][]{FWilczek_QCDaxion, SWeinberg_QCDaxion} were experimentally ruled out by several laboratory experiments \citep[][]{FaissnerProtonBeam, NuclearDeExcitation,NegativeAxionsResultsZenhder,ZEHNDERRuledOutStandardAxion1982}.

Axion-Like Particles (ALPs) are a model-independent generalisation of the QCD axion, the mass and coupling to electromagnetism of which need {not} be proportional to each other, and are predicted by several BSM extensions including string theory \citep[see, e.g,][]{greenSchwarzWitten_2012}. In the early Universe, the dominant production channels of ALPs of masses $\lesssim 10 ^{-6} \ \mathrm{eV}$ will be non-thermal. Interestingly, ALPs could comprise Dark Matter (DM) \citep[][]{PRESKILL_Wise_Wilczek_CosmoInvisibleAxion_83, Abbott_Sikivie_CosmologicalBoundOnAxion_83,Dine_Fischler_83_HarmlessAxion}. On the other hand, \textit{very light} ALPs of masses $< 10 ^{-33} \ \mathrm{eV}$ could contribute to the Dark Energy component of the Universe \citep[][]{ALPsQuintessentialDE_Carroll}. Thereby, probing ALPs probes the validity of these BSM theories.

The goal of this paper is to set an upper bound on the coupling of light ALPs to electromagnetism, $g_\mathrm{a\gamma}$, as a function of the ALP rest mass, $m_\mathrm{a}$, independent of their role in cosmology. In the presence of an external magnetic field \textbf{B}, the strength of the axion/two-photon interaction (thereby, equivalent for ALPs) is described by the interaction Lagrangian:\begin{equation}
    \label{eq: Lagrangian_vertex}
    \Centering
    \mathcal{L}_\mathrm{a\gamma} = g_\mathrm{a \gamma}\,a\,\textbf{B} \cdot \textbf{E},
    \Centering
\end{equation}
where \textit{\textbf{E}} and $a$ are the electric and the ALP fields, respectively. For a photon beam propagating through a background magnetic field that has some non-zero component perpendicular to the direction of propagation, $\mathcal{L}_\mathrm{a\gamma}$ describes the mixing of the photon and ALP states. We note that, at energies below the Fermi scale, all ALP interactions with fundamental SM particles through the weak interaction can be disregarded, since $\mathcal{L}_\mathrm{a\gamma}$ dominates. 

Eq. \ref{eq: Lagrangian_vertex} describes the low-energy phenomenology of ALPs, which can be explored with a variety of laboratory and astrophysical systems. In particular, astrophysical objects emitting in the $X-\gamma$ ray bands provide a powerful tool to probing \textit{light} ALPs. For photons of energy $E \gg m_\mathrm{a}$ travelling through a fully ionised plasma of uniform electron number density $n_\mathrm{e}$, one can derive a photon-beam propagation equation incorporating the interaction described by Eq. \ref{eq: Lagrangian_vertex}. Using the short-wavelength approximation, the latter becomes a Schrodinger-like Equation \citep[][]{Raffelt:1987_MixingOfPhoton}. 

In a simplified geometry with a homogeneous background magnetic field, the Schrodinger-like Equation becomes an eigenvalue problem. In the $X-\gamma$ bands, one can ignore the sub-leading effect of Faraday Rotation, since it scales as $\sim E^{-2}$. In this regime, the characteristic or mixing matrix of the eigenvalue problem induces three phenomena: photon-ALP mixing; photon-ALP resonant inter-conversion; and the usual energy-dependent dispersion of photons propagating through a plasma.

For a given ALP mass $m_\mathrm{a}$ and coupling strength $g_\mathrm{a\gamma}$, the eigenvalue problem yields the probability $P_\mathrm{a\gamma}$ that an initial polarisation beam state will transform into an ALP state after having travelled distance $L$ along the line-of-sight, and vice-versa. If in a suitable location of parameter space, an energy-dependent photon-ALP mixing
probability $P_\mathrm{a\gamma}$ exhibits energy-dependent oscillatory features (e.g. see Fig. \ref{figure:MixingCurves}) at a characteristic
scale $E(g_\mathrm{a\gamma}) \times g_\mathrm{a\gamma} \ll 1$. Generically, for non-simplified field geometries, $P_\mathrm{a\gamma}$ is a non-trivial function that depends on $E$, $L$, $\textbf{B}$, and on $m_\mathrm{a}$, $g_\mathrm{a\gamma}$.

Stars were the first astrophysical objects used to probe light ALPs (of $m_\mathrm{a}\lesssim 10^{-10} \ \mathrm{eV}$), since the scattering of ionised particles and the high radiation density in stellar cores would produce ALPs if $g_\mathrm{a\gamma} \neq 0$ \citep[][]{Raffelt:1996_StarsAsLabsForAxions, Raffelt07_reviewPrimakkoff}.

Historically, the earliest limit on ALPs for an extended mass range ($m_\mathrm{a} < 0.03 \ \mathrm{eV}$) came from the analyses of the lifetime of helium burning stars in globular clusters \citep[for recent constraints, see][]{PierlucaCarenza_HeliumBurningStars_MostRecent, HeliumBurningStars_Constraints_2}. In addition, the CERN Axion Solar Telescope (CAST) searched for ALPs generated at the core of the Sun by re-processing them into X-ray photons within the experiment. The lack of evidence for such ALPs at CAST has excluded $g_\mathrm{a\gamma} > 6.6\times 10^{-11} \ {\mathrm{GeV}}^{-1}$ for $m_\mathrm{a} \lesssim 0.02 \ \mathrm{eV}$ \citep[][]{CAST_AbsenceOfALPs}. 

\begin{table}
\centering 
\renewcommand{\arraystretch}{1.2}
\begin{tabular}{c c c}
\toprule 
\midrule
    AGN & Upper bound on $g_\mathrm{a\gamma}[\mathrm{GeV}^{-1}]$ \\ \hline \hline
    Hydra A & $8.3 \times 10^{-12}$ for $m_\mathrm{a} < 7 \times 10^{-12} \ \mathrm{eV}$ \\ \hline
    M87 & $2.6 \times 10^{-12}$ for $m_\mathrm{a} \leq 1 \times 10^{-13}\ \mathrm{eV}$\\ \hline
    NCG3862 & $2.4 \times 10^{-12}$ for $m_\mathrm{a} \lesssim 10^{-12}\ \mathrm{eV}$\\ \hline
    2E3140 & $1.5 \times 10^{-12}$ for $m_\mathrm{a} \lesssim 10^{-12}\ \mathrm{eV}$\\ \hline
    \multirow{2}{*}{NCG1275} & \multicolumn{1}{c}{$(1.4 - 4.0) \times 10^{-12} \ {\mathrm{GeV}}^{-1}$ for $m_\mathrm{a} \lesssim 10^{-12} \ \mathrm{eV}$} \\ 
\cmidrule{2-2}
  & $8\times 10^{-13}$ for $m_\mathrm{a} < 1 \times 10^{-12} \ \mathrm{eV}$ \\ \hline
    H1821$+$643 (Our work) & $6.3 \times 10^{-13}$ for $m_\mathrm{a} < 10^{-12} \ \mathrm{eV}$\\
\bottomrule
\end{tabular}
\caption{\label{table:Tab1_Constraints_centralAGN_CoolCore}Constraints inferred from the lack of spectral distortions of AGN spectra, in order of appearance, by: \citet[][]{Wouters_FirstConstraints}, \citet[][]{Marsh17}, \citet[][]{Conlon_17_manySources}, \citet[][]{Conlon_17_manySources}, \citet[][]{Berg:2016_NGC1975}, Model B of \citet[][]{Reynolds20}, and our work (quoted at $95\%$, $95\%$, $95\%$, $95\%$, $99.7\%$ and $99.7\%$ credibility). We note that throughout our paper, when quoting \citet[][]{Reynolds20}, we \textit{only} refer to the results from Model B of \citet[][]{Reynolds20}, since they are the most comparable to ours.}\end{table}

Similarly, the absence of a gamma-ray signal linked to the re-processing of ALPs generated at the core of SN1987A set the upper bound $g_\mathrm{a\gamma} \lesssim 5.3 \times 10^{-12} \ {\mathrm{GeV}}^{-1}$ for $m_\mathrm{a} \lesssim 4.4 \times 10^{-10} \ \mathrm{eV}$ \citep[][]{SN1987Bounds}. Recently, a hard \textit{NuSTAR} observation of Betelgeuse set $g_\mathrm{a\gamma} < (0.5 - 1.8) \times 10^{-11} \ {\mathrm{GeV}}^{-1}$ for $m_\mathrm{a} < (3.5 - 5.5) \times 10^{-11} \ \mathrm{eV}$ \citep[][]{Betelegeuse_21}, on the basis of similar magnetic field assumptions.

Key to our work lies the notion that rich clusters of galaxies hosting bright Active Galactic Nuclei (AGN) are efficient photon-ALP inter-converters. All clusters are permeated by a magnetised plasma of typical field intensities of $B \sim0.1 - 1.0 \ \mathrm{\mu G}$, namely, the intracluster medium (ICM). Their suitability to constrain ALPs is encapsulated by $B \times L$, where $L \sim0.1 - 1.0 \ \mathrm{Mpc}$ is the distance travelled by AGN photons through the cluster line-of-sight. In clusters, this product can greatly exceed values attainable in laboratory experiments. If in a suitable location of parameter space, a non-vanishing $P_\mathrm{a\gamma}$ would effectively mimic absorption features in the spectrum of an embedded AGN collected in the cluster vicinity. Therefore, given an appropriate model for the cluster magnetic field, the absence of such spectral distortions in the X-ray spectrum of the AGN can be used to constrain the photon-ALP coupling.

Table \ref{table:Tab1_Constraints_centralAGN_CoolCore} lists the best constraints on light ALPs that have been extracted by independent detailed spectroscopic \textit{Chandra} analyses of six cluster-embedded AGN, including this work. Firstly, \citet[][]{Wouters_FirstConstraints} used imaging of the faint AGN at the Hydra-A Brightest Cluster Galaxy (BCG) to constrain $g_\mathrm{a\gamma}$ and already exceeded the capabilities of concurrent laboratory experiments \citep[][]{Graham15_LightThroughWall}. Nevertheless, the separation between the intrinsic AGN and cluster contributions was limited by the strong astrophysical absorption present within $1 - 7 \ \mathrm{keV}$. Their results were tightened by \citet[][]{Marsh17}, who employed short ($0.4 \ \mathrm{s}$) frame time observations of the extraordinarily bright embedded AGN within Virgo, M87. Concurrently, \citet[][]{Conlon_17_manySources} employed archival \textit{Chandra} observations of seven AGN hosted by or located behind clusters to constrain ALPs of $m_\mathrm{a} \lesssim 10^{-12} \ \mathrm{eV}$. In their sample, the most promising sources for ALP studies were identified to be NCG3862 in the A1367 cluster and the Seyfert galaxy 2E3140 in A1795. Moreover, \citet[][]{Berg:2016_NGC1975} used an extended set of short archival \textit{Chandra} and \textit{Newton-XMM} observations of the central engine of the Perseus cluster, NGC1275, to exclude $g_\mathrm{a\gamma} > (1.4 - 4.0) \times 10^{-12} \ {\mathrm{GeV}}^{-1}$ depending on the field model along the line-of-sight. Pile-up was mitigated by the use of \textit{Chandra} observations where the AGN had been located midway and on-edge relative to the focal point. Most recently, \citet[][]{Reynolds20} employed a deep ($490 \ \mathrm{ks}$) \textit{Chandra}-HETG (High Energy Transmission Grating) observation of NGC1275 to infer the tightest constraints that had been set at the time on ALPs of $\mathrm{log}(m_\mathrm{a}/\mathrm{eV}) < -12.0$. The use of the HETG enabled the extraction of an AGN spectrum free from pile-up. However, the bright emission of the ICM complicated the treatment of the background subtraction, leaving instrumental residuals at the $3\%$ level. Most recently, \citet[][]{Schallmoser2021_updated_ML} have reanalysed the observations employed by \citet[][]{Conlon_17_manySources} using machine learning techniques to derive the improved constraint of $g_\mathrm{a\gamma} \lesssim 6.0 \times 10^{-13} \ {\mathrm{GeV}}^{-1}$ at $95\%$ credibility.

In our work, we employ a combined Low-Energy and High-Energy Transmission Spectroscopy Grating (thereby, LETG and HETG) \textit{Chandra} observation of the radio-quiet quasar H1821+643 to infer the tightest constraints to-date on light ALPs. This is an extraordinarily luminous quasar, $L_\mathrm{bol} \sim 2 \times 10^{47} \mathrm{erg/s}$, located at the centre of the BCG of the cool-core cluster CL1821+643 \citep[][]{Russell10}. This cluster is believed to host one of the most massive X-ray bright Super Massive Black Holes (SMBHs) found to date at a cosmologically interesting distance ($d \sim 1 \ \mathrm{Gpc}$). Indeed, X-ray observations of this cluster-AGN system have suggested that the density of the ICM at the BCG can be up to twice as large as that in Perseus \citep[][]{Russell10, DeepChandra_Perseus_Fabian06}. Using a total on-source exposure of $571 \ \mathrm{ks}$ of this remarkable system together with a fiducial magnetic field model, we exclude all couplings $g_\mathrm{a\gamma} > 6.3 \times 10^{-13} \ {\mathrm{GeV}}^{-1}$ for most ALPs of masses $m_\mathrm{a} < 10^{-12} \ \mathrm{eV}$ at $99.7\%$ credibility. At $95.5\%$ confidence, we exclude $g_\mathrm{a\gamma} > 5.0\times 10^{-13} \ {\mathrm{GeV}}^{-1}$ within the same mass range. 

In the following section, we present the archival spectroscopic \textit{Chandra} observations of H1821+643 we have employed, as well as our fiducial quasar model. We also introduce our best fits to the thermal properties of the host cluster. In Sec. \ref{sec:s3_ALPSurvivalCurves}, we describe how we have generated a grid of photon-ALP mixing models we fit to the quasar spectrum to infer posteriors on ALP parameters following the Bayesian technique described in Sec. \ref{sec:s4_toInferPosteriors}. We present our results and conclusions in Sections \ref{sec:s5_Results_Discussion} and \ref{sec:s5_Conclusions}, respectively, and explore the sensitivity to several assumptions we have made in the Appendix. 

\section{Observations, data reduction and fiducial quasar model}
\label{sec2:Obs_and_dataRed}
H1821+643 is a powerful radio-quiet quasar that has been the subject of intense study, particularly in the search of absorption lines from the Warm-Hot Intergalactic Medium \citep[][]{Fang_2002_WHIMObs_h1821+643,Mathur_2003_WHIM_h1821+643,Kovacs_LETGChandraObs_MissingBaryons}. This quasar resides at the centre of a massive cool-core cluster of galaxies, CL1821+643, whose ICM temperature and entropy profiles steepen significantly at $\sim 80 \ \mathrm{kpc}$ to the cluster centre \citep[][]{Russell10}. The upper bound on the mass of the SMBH at the cluster centre, $\mathrm{log}(M_\mathrm{SMBH}/M_\odot) \leq 10.5$, was inferred from an argument where a Compton cooling cycle were to uniquely explain the (rare) low-entropy cluster core. Additionally, X-ray reflection spectroscopy reported a lower mass bound of $\mathrm{log}(M_\mathrm{SMBH}/M_\odot) \geq 9.5$ \citep[][]{Reynolds14}, also providing a tentative spin constraint for the central SMBH ($a > 0.4$ in dimensionless units). This corresponds to the only spin constraint that has been found to date for such a massive accreting system \citep[see Fig. 6 of][]{reynolds2019ObsBlackHoleSpinsReview}. 

This source is located at redshift $z = 0.299$ \citep[][]{ArticleComputingZ_0.299_ForH1821+643}. We assume a flat ($\Omega_\mathrm{\kappa} = 0.0$) Universe compatible with $\Lambda$CDM cosmology, that is, assuming: $H_0 = 70 \ \mathrm{km/s/Mpc}$, $\Omega_\mathrm{m} = 0.3$, $\Omega_\mathrm{\Lambda} = 0.7$ \citep[][]{PlanckCollaborationLatestResults18}, which translates into a luminosity distance of $1.55 \ \mathrm{Gpc}$. We assume a hydrogen column density local to the Milky Way of $N_\mathrm{H} = 3.51 \times 10^{20} \ \mathrm{cm^{-2}}$ \citep[][]{Reference_forNH}. We take the element abundance ratios of \citet[][]{1989_XspecAng_abundances}. 

\subsection{Transmission Grating observations of H1821+643}
\label{sec2.1:h1821+643_ChandraSpectroscopyData}

A low-energy transmission grating (LETG) \textit{Chandra} observation of H1821+643 took place over 2001 January 17 - 24. It used the \textit{Chandra}'s Advanced CCD Imaging Spectrometer (ACIS) S array and a total exposure of $471.4 \ \mathrm{ks}$, taking place in four segments (Obs IDs - exposure: 2186 - $165.4 \ \mathrm{ks}$; 2310 - $163.7 \ \mathrm{ks}$; 2311 - $90.5 \ \mathrm{ks}$; and 2418 - $51.8 \ \mathrm{ks}$). We also include the only high-energy transmission grating (HETG) observation of the quasar (Obs ID: 1599 - $99.6 \ \mathrm{ks}$), also on ACIS-S, which started on 2001 February 9. The HETG consists of the High-Energy and Medium Energy Gratings (HEG and MEG, respectively).

\begin{figure}
 \center
 \includegraphics[width=0.5\textwidth]{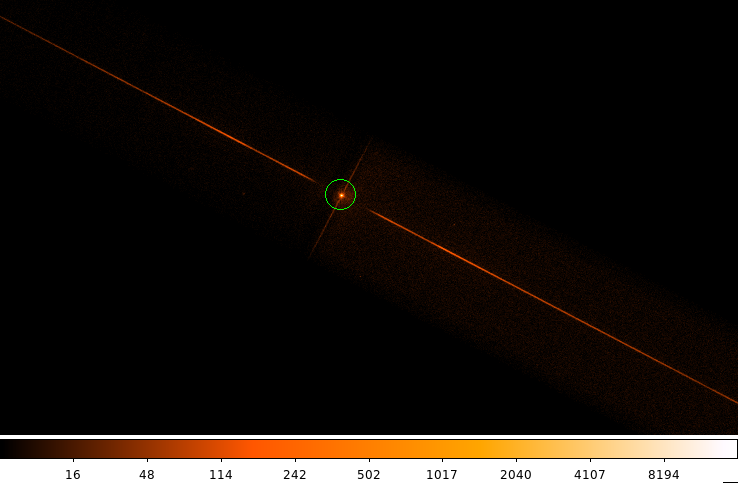}
 \caption{Full-band image of ACIS-S array for ObsID 2186 (LETG; exposure $165.4 \ \mathrm{ks}$). For display, the raw pixel data is binned by 2$\times$2, and the colour bar shows total photon count per (new) binned pixel. The green circle indicates the source extraction region. Although the gratings are slitless and hence the cluster emission is also dispersed, the AGN component dominates.}
 \label{figure:DispersedQuasarSpectra_HETG_Reynolds20}
\end{figure}

Figure \ref{figure:DispersedQuasarSpectra_HETG_Reynolds20} shows the image of the ACIS-S array for LETG observation ObsID 2186, the observation with the single longest exposure. Although both the LETG and HETG instruments consist of slitless gratings which will disperse both the cluster light and AGN emission, the AGN emission is dominant and can be well isolated from the ICM. Additionally, order-sorting only accepts photons with CCD-detected energies compatible with their spatial position along the dispersion spectrum. This further serves to isolate the AGN spectrum.

The archival \textit{Chandra} observations were reprocessed with CIAO-4.13 and CALDBv4.9.4. We follow standard CXC science threads\footnote{\url{https://cxc.cfa.harvard.edu/ciao/threads/spectra_hetgacis/} and \url{https://cxc.cfa.harvard.edu/ciao/threads/spectra_letghrcs/}} for the extraction of the AGN spectra for each observation with one exception. For the HETG observation only, the width of the spectral extraction region was reduced by approximately half (\texttt{width\_factor\_hetg=18}) which reduces overlap of the MEG and HEG at the centre of the array thus improving the range of the HEG at higher energies. For the LETG observations, a time-averaged spectrum was produced using CIAO's \texttt{combine\_grating\_spectra}. 

We perform a joint analysis of these reduced HETG/LETG spectra to infer a high-resolution spectrum of H1821+643, free from pile-up, within $0.8 - 8.5 \ \mathrm{keV}$ (observed energies). Individually, each grating accounts for the following observed energies: $1.5 - 8.5 \ \mathrm{keV}$ (HEG), $1.0 - 7.0 \ \mathrm{keV}$ (MEG), and $0.8 - 6.5 \ \mathrm{keV}$ (LEG). Our analysis has produced the highest quality spectrum of H1821+643 to date.
\subsection{Fiducial quasar model for H1821+643}
\label{sec:s2.2_FiducialQuasarMod_h1821+643} Throughout our spectral quasar analysis, we have employed version 12.11.1 of the \textsc{xspec} X-Ray Spectral Fitting Package \citep[][]{Xspec_LatestRef_1996Arnaud}. We infer a best-fit astrophysical model for the quasar spectrum based on the minimisation on the \textit{Cash} (\textit{C}) statistic of the combined fit, to which each individual dataset contributes linearly. The \textit{C}-stat of a fit quantifies the goodness-of-fit attributed to a model applied to describe Poisson-distributed data \citep[][]{CashStatisticsUse_Kaastra_Astronomers}. 

\begin{table}
\centering 
\renewcommand{\arraystretch}{1.2}
\begin{tabular}{c c c c c}
\toprule 
\midrule
    Model & HETG & LETG \\
    \hline \hline 
    \multirow{2}{*}{\textsc{PO[A]}} & $\Gamma_\mathrm{0} = 2.900 \pm 0.921 $ & 
    $\Gamma_\mathrm{0} = 2.798 \pm 0.294 $ \\
    & $A_{0} = (2.044 \pm 0.002)\times {10}^{-3}$ & $A_{0} = (2.484 \pm 0.001)\times {10}^{-3}$ \\
    \hline
    \multirow{2}{*}{\textsc{diskline}} & $A_{1} = (5.130 \pm 1.191)\times {10}^{-5}$ & $A_{1} = (1.907 \pm 0.535)\times {10}^{-5}$ \\ & $\mathrm{EW}_\mathrm{1} = 0.308 \ \mathrm{keV}$ & 
    $\mathrm{EW}_\mathrm{1} = 0.124 \ \mathrm{keV}$ \\ 
    \hline
    \multirow{2}{*}{\textsc{zgau}} & $A_{2} = (2.312 \pm 1.368)\times {10}^{-3}$ & $A_{2} = (2.484 \pm 0.876)\times {10}^{-3}$ \\ & $\mathrm{EW}_\mathrm{2} = 0.031 \ \mathrm{keV}$ & 
    $\mathrm{EW}_\mathrm{2} = 0.028 \ \mathrm{keV}$ \\ \hline
    \multirow{2}{*}{\textsc{PO[B]}} & $\Gamma_\mathrm{3} = 1.611 \pm 0.290 $ & $\Gamma_\mathrm{3} = 1.644 \pm 0.161 $ \\
    & $A_{3} = (2.312 \pm 0.002)\times {10}^{-3}$ & 
    
    $A_{3} = (2.178 \pm 0.001)\times {10}^{-3}$
    \\ 
\bottomrule
\end{tabular}
\caption{\label{table:Details_bestFitParValues} Best-fit values of the free parameters that are left free in the fiducial model employed in our HETG/LETG combined analysis of H1821+643, where the \textsc{diskline} inclination (\textit{Inc}) free parameter is given by $43.083^\circ \pm 0.850^\circ$. We show the equivalent widths ($\mathrm{EW}$) of the \textsc{diskline} and \textsc{zgau} model parameters for both the HETG (averaged over HEG/MEG) and LETG datasets. All normalisation parameters $A_\mathrm{i}$ are quoted in units of $\mathrm{counts/s/keV}$.}
\end{table} Throughout our spectral analysis, we exclude all photon events with observed energies below $0.8 \ \mathrm{keV}$, given the strong noise-like features present in the LEG dataset below this threshold. We find that the intrinsic AGN emission is well described by a double power law, \textsc{PO(A) $+$ PO(B)}. Additionally, we consider the effects of Galactic absorption in both the continuum and reflection components via the Tuebingen-Boulder model \textsc{TBABS} \citep[][]{Wilms_2000}, with a constant hydrogen column density of $N_\mathrm{H} = 3.51 \times 10^{20} \ \mathrm{cm}^{-2}$ across all datasets. We also include a multiplicative constant \textsc{CONST} with the purpose of correcting for potential calibration uncertainties associated with each grating.

Our choice of spectral model for the embedded AGN was motivated as follows. We adopt the \textsc{diskline} model to describe the relativistic \textsc{Fe}K$\alpha$ emission associated with the coronal reflection from the inner accretion disc \citep[][]{Fabian1989_Diskline}. Power-law models are ubiquitous in AGN X-ray spectra. We initially fitted a single power-law continuum, yielding a best-fit spectral index of $1.937 \pm 0.018$ (HETG) and $2.082 \pm 0.006$ (LETG), and normalisation parameters of $(4.036 \pm 0.018) \times 10^{-3} \ \mathrm{counts/s/keV}$ (HETG), and $(4.548 \pm 0.019) \times 10^{-3} \ \mathrm{counts/s/keV}$ (LETG). To start with, we used the \textsc{diskline} approximation for the relativistically broadened iron line at an unconstrained inclination. Nevertheless, fitting a single power-law model with such parameters results in a soft excess. Our description of the intrinsic AGN emission improves by fitting a double power law continuum (\textsc{PO[A]} and \textsc{PO[B]}, with best-fit parameter values listed in Tab. \ref{table:Details_bestFitParValues}) by a factor of $\Delta C = -206.3$ overall. However, a double power-law fit leaves remaining structure in the iron band, for which we include a redshifted gaussian \textsc{zgau} in our model, to account for a distant, cold reflector. This improves the fit by $\Delta C = -42.3$ globally.

We fit the \textsc{CONST*TBABS*(PO[A]+DISKLINE+ZGAU+PO[B])} model to the combined HETG/LETG observation of H1821+643. The inferred best-fit values of this model, shown in Table \ref{table:Details_bestFitParValues}, define our fiducial or baseline model for the quasar spectrum. We note that, given the employed HETG/LETG observations were taken at different times, we let our model account for potential AGN spectral time variability by untying all free parameter models that quantify the intrinsic and reflected AGN components attributed to different observation times. In particular, the former are: each of the two power laws' (\textsc{PO[A]} and \textsc{PO[B]}) spectral indices, $\Gamma_\mathrm{i}$, and their corresponding normalisations, $A_\mathrm{i}$. This particularly characterised the spectral modelling of the HETG data, since the HEG \textsc{constant} model component was frozen at 1.0, whereas that associated with the MEG was instead set free, which led to a best-fit parameter value of $1.004 \pm 0.065$. This enabled the consideration of different calibration uncertainties attributed to each of the two gratings mounted on \textit{Chandra}'s HETG. 

\begin{figure*}
 \centering
  \includegraphics[width=.85\textwidth]{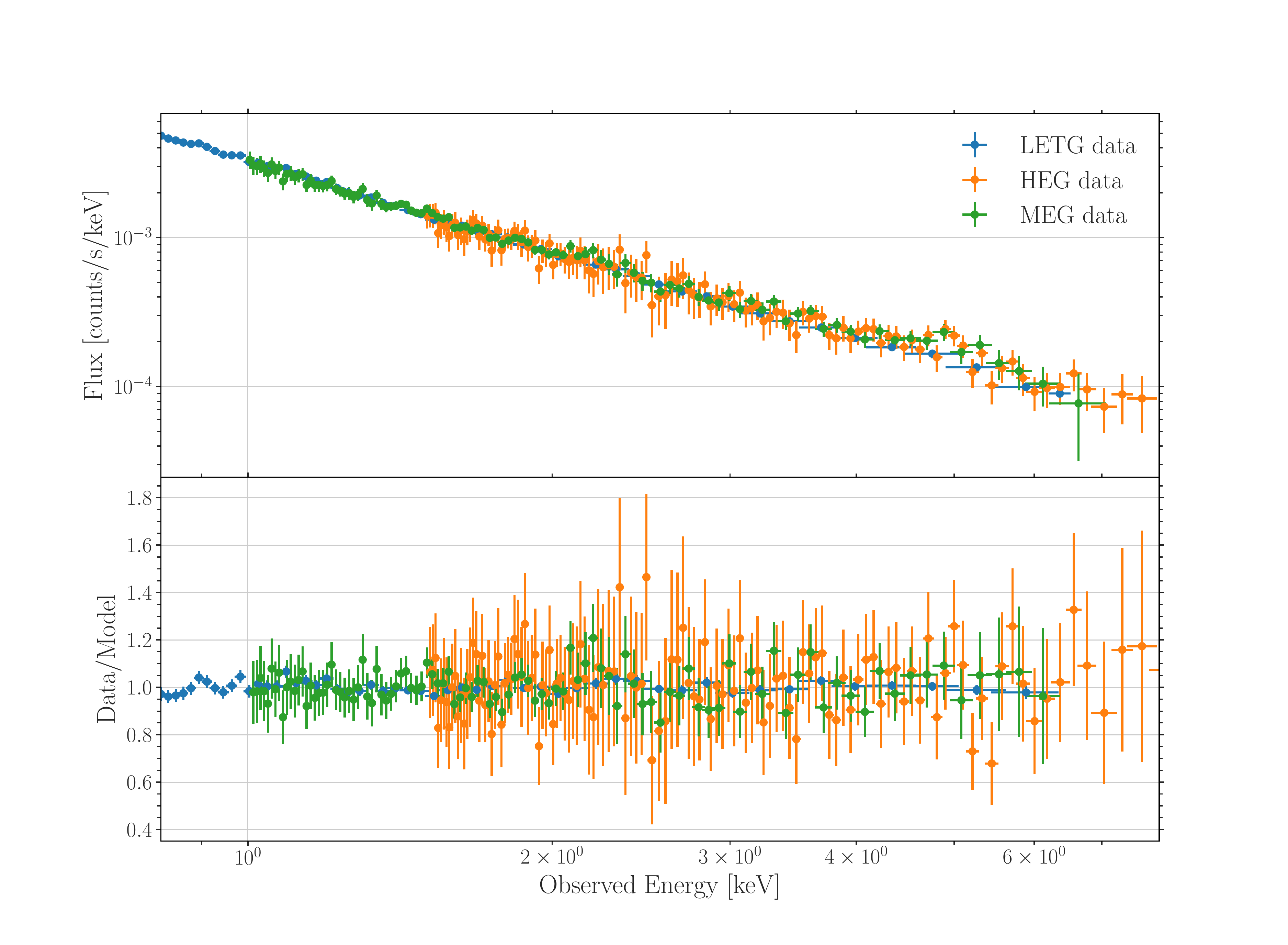}
  \caption{Data (upper panel) and ratio of the data to the best-fit model to the observations of H1821+643 employed in our analysis. All error bars represent uncertainties within $1\sigma$. All spectral data have been binned so as to achieve a $S/N \geq 200$, under the condition that 20 spectral bins at maximum can be merged. We note that this binning scheme does not affect our ALP constraints, since our ALP fitting scripts were performed onto the unbinned spectral data.}
  \label{figure:myModFiducial}
\end{figure*}

As seen in Fig. \ref{figure:myModFiducial}, the quasar spectrum presents spectral features at $6 - 7 \ \mathrm{keV}$ (rest energies). To account for spectral X-ray variability, we enable \textsc{diskline}'s \textit{normalisation} parameter to vary amongst the HETG and LETG datasets, assuming an inner radius of $R_\mathrm{in} = 6$ (in units of gravitational radius, $R_\mathrm{g}$) for both datasets. The latter corresponds to the innermost stable circular orbit (ISCO) for a non-rotating black hole \citep[i.e. of spin $a = 0$, see Sec. 2 of][]{reynolds2019ObsBlackHoleSpinsReview}. Given that our aim is \textit{not} to perform spin studies of H1821+643, we can expect this to be a reasonable assumption. Furthermore, consistent with a default coronal lamppost geometry of height $h[R_\mathrm{g}] = 6$ above the centre of the SMBH, we assume an emissivity index of $\xi = 3$, and an outer disc truncation radius of $R_\mathrm{out}[R_\mathrm{g}] = 10^{3}$. 

We note that fitting a double power law continuum results in degeneracies between the free parameters associated with both the power law (\textsc{PO[A]}, \textsc{PO[B]}) and the \textsc{diskline} models. This is suggested by the significantly larger error attributed to the HETG best-fit value of the photon index of the \textsc{PO[A]} model component ($\Gamma_0$ in Tab. \ref{table:Details_bestFitParValues}). To prevent our spectral fit from getting stuck at a local minimum when attempting to conduct an ALP search, we froze the disc inclination \textit{Inc} to its best-fit value, $43^{\circ}$. Although, ideally, we would have wanted this unknown parameter characterising the orientation of accretion disc to be free, our approach is valid as long as the treatment of the \textsc{diskline}'s \textit{Inc} free parameter is also frozen at $43^{\circ}$ for all spectral models containing ALPs (see Sec. \ref{sec:s3_ALPSurvivalCurves}). 

\begin{table}
\centering 
\renewcommand{\arraystretch}{1.2}
\begin{tabular}{c c c c c}
\toprule 
\midrule
    Magnitude & HEG & MEG & LEG \\ \hline \hline
    $L [\mathrm{erg}/ \mathrm{s}]$ & $3.76\times 10^{45}$ & $4.18\times 10^{45}$ & $4.15\times 10^{45}$ \\ \hline
    $F [\mathrm{erg/{cm}^{2}/ {s}}]$ & $1.79 \times 10^{-11}$ & $1.52\times 10^{-11}$ & $1.47\times 10^{-11}$ \\ \hline
    \textit{C}-stat & $2260.40$ & $2023.14$ & $1206.68$ \\ 
    DOF & $2454$ & $2098$ & $1080$ \\
\bottomrule
\end{tabular}
\caption{\label{table:FinalDetailsFiducialModel} Unabsorbed band luminosity ($L$) and flux ($F$) computed over $0.8 - 8.5 \ \mathrm{keV}$ (observed energies); and $C$-stat and degrees of freedom (DOF) attributed to the HEG, MEG and LEG datasets employed in our study.}
\end{table}

Our high quality spectrum of H1821+643 free from photon pileup, yields: a combined unabsorbed X-ray band luminosity and flux of $L_\mathrm{0.8 - 8.5 \ keV} = 1.21 \times 10^{46} \ \mathrm{erg}/\mathrm{s}$, $F_\mathrm{0.8 - 8.5 \ keV} = 4.78 \times 10^{-11} \ \mathrm{erg}/\mathrm{cm}^{2}/\mathrm{s}$ (observed energies); and a \textit{C}-stat = 5490.22 for 5632 Degrees of Freedom (DOF). The individual contributions to these magnitudes from each of the three gratings are listed in Tab. \ref{table:FinalDetailsFiducialModel}. Our baseline quasar model results in spectral distortions $< 2.5\%$ (see Fig. \ref{figure:myModFiducial}). 
\subsection{Fits to thermal profiles of the host cluster}
\label{sec:s2.3_Russell10_descriptionAndTreatment}
We now proceed to describe our model for the thermal properties of the host cluster relevant to our ALP search, namely its thermal pressure $P_\mathrm{th}(r)$ and electron number density $n_\mathrm{e}(r)$ profiles (see Fig. \ref{figure:VerticallyStackedFig_H1821_profiles}). We use the former to estimate the magnetic field intensity throughout the cluster volume, shown by Fig. \ref{eq:BFieldIntensityAtClusterCenter}, assuming that the ratio of thermal-to-magnetic pressure $\beta$ remains constant up to the virial radius. On the other hand, $n_\mathrm{e}(r)$ determines the plasma frequency distribution of the ICM. In Sec. \ref{sec:s3_ALPSurvivalCurves}, we will outline how we have employed this magnetic field model to generate a grid of photon-ALP mixing curves we then fit to the combined HETG/LETG quasar spectrum and compare to our baseline quasar spectral model to constrain ALPs. 

In 2010, \textit{Chandra} observed CL1821+643 in four sets of imaging observations on ACIS-S (Obs IDs - exposure time: 9398 - $34 \ \mathrm{ks}$, 9845 - $25 \ \mathrm{ks}$; 9846 - $18 \ \mathrm{ks}$; 9848 - $11 \ \mathrm{ks}$), with a resulting cleaned exposure of $85 \ \mathrm{ks}$. These grating-free data were analysed by \citet[][]{Russell10} using the processing and analysing softwares \textsc{CIAO 4.1.2} and \textsc{CALDB 4.1.2}. The image region within the inner $3 ''$ ($r\leq 10 \ \mathrm{kpc}$) was excluded due to the strong pile-up affecting the AGN emission.

In \citet[][]{Russell10}, the authors employed the Direct X-ray Spectral Deprojection method \citep[\textsc{DSDEPROJ}, see][]{2008_DSDEPROJ_Russel, SandersAndFabian_2007_firstDirectXRaySpatialDeproj} on their processed \textit{Chandra} imaging observations of CL1821+643 to map its thermal profiles. \textsc{DSDEPROJ} relies on the assumption of cluster spherical symmetry, and produces a set of deprojected spectra based on the subtraction of successive outermost annuli down to the cluster centre.

Our best fit to the cluster thermal pressure $P_\mathrm{th}(r)$ and the electron number density $n_\mathrm{e}(r)$ of the \textsc{DSDEPROJ} data of \citet[][]{Russell10} is shown by Fig. \ref{figure:VerticallyStackedFig_H1821_profiles}. We fit a bending power law (PL) to $P_\mathrm{th}(r)$ falling at $R_\mathrm{b}[\mathrm{kpc}]$ to the cluster centre, via:
\begin{equation}
\centering
\label{eq:h1821_PRESSUREprofileFits} P_\mathrm{th}(r) = P_0 \frac{r^{-\gamma_1}}{1 + {(r/R_\mathrm{b})}^{\gamma_2 - \gamma_1}},
\end{equation}
finding the following best-fit parameter values: $P_0 = (1.85 \pm 0.07) \times 10^{-9} \ \mathrm{erg} / \mathrm{cm}^{3}$, $R_\mathrm{b} = 511 \pm 18 \ \mathrm{kpc}$, $\gamma_1 = 0.47 \pm 0.01$, and $\gamma_2 = 2.54 \pm 0.02$. We fit the number density profile $n_\mathrm{e}(r)$ with a decaying PL of amplitude $2.63 \pm 0.15 \ \mathrm{cm}^{-3}$ and index $1.16 \pm 0.01$. We refer the reader to Appendix \ref{sec:sec6.1_thermalPropertiesOfHostCluster} for further details on our fits, and on how one may employ them to estimate $M_{500}$ -- the mass of a spherical object within which the mean density is $500$ times the critical density of the Universe --, which is commonly used as a proxy for the cluster mass.
 
\section{Generating a grid of ALP survival curves}
\label{sec:s3_ALPSurvivalCurves}
We now outline our computation of an extended library of photon-ALP mixing curves which we fit to the combined HETG/LETG spectrum of H1821+643 to constrain ALP parameter space. Following Model A of \citet[][]{Reynolds20}, and given the absence of thermal pressure data within $r\leq 10 \ \mathrm{kpc}$ to the cluster centre (see Fig. \ref{figure:VerticallyStackedFig_H1821_profiles}), we conservatively omit all photon-ALP inter-conversion taking place within this region. We therefore only consider photon-ALP mixing occurring to photons emitted by the AGN as they travel through the magnetised ICM within $r = 10 \ \mathrm{kpc} - 1.5 \ \mathrm{Mpc}$ to the cluster centre. 

We would expect this upper bound ($1.5 \ \rm{Mpc}$) to be located within the virial radius of the cluster, and thereby, to be moderately conservative. We refer to \citet[][]{Russell10}, where the surface brightness deprojection method predicted an overdensity radius of $R_{200} = 2.5^{+1.3}_{-0.7}\ \mathrm{Mpc}$. To a first order, $R_{200}$ serves as a proxy for the virial radius of the cluster \citep[][]{1998ApJ_BryanAndNorman_R200_andRVir_estimates}.
\begin{figure}
 \center
  \includegraphics[width=0.45\textwidth]{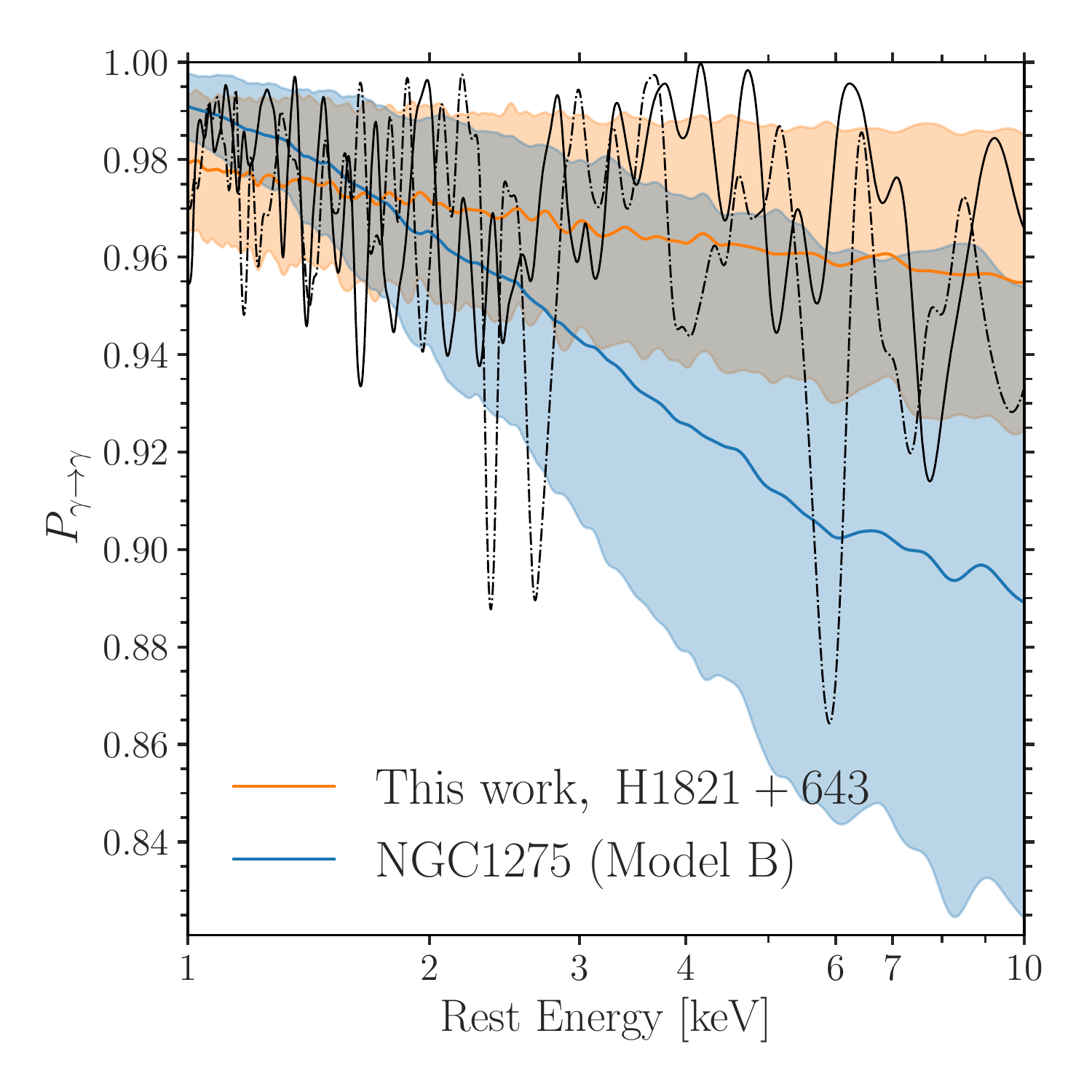}
  \caption{Mean photon survival probability as a function of energy for our H1821$+$643 model (orange) and NGC1275 \citep[blue, Model B of][]{Reynolds20}. The shaded regions show the standard deviations of 500 curves in each energy bin, computed by using 1000 logarithmic bins between $1 - 10 \ \rm{keV}$. The black solid and dot-dashed lines show survival probability curves for two individual realisations of the magnetic field we have produced. All calculations are performed for ALP parameters $g_\mathrm{a\gamma}=10^{-12}\,\mathrm{GeV}^{-1}$, $m_\mathrm{a}=10^{-13} \rm{eV}$.}
  \label{figure:MixingCurves}
\end{figure}

At a given location within the cluster, of co-moving distance $r$ to the cluster centre, the ICM plasma frequency is, in energy units: \begin{equation}
\centering
\label{eq:PlasmaFrequency}
     \omega_\mathrm{pl}(r) = 4.2 \times 10^{-12} \times \biggl({\frac{r}{100 \ \mathrm{kpc}}}\biggl)^{-0.58} \ \mathrm{eV},
\end{equation} as derived from our best fit to the \citet[][]{Russell10} \textsc{DSDEPROJ} electron density profile, $n_\mathrm{e}(r)$. For all ALP masses $m_\mathrm{a} > \omega_\mathrm{pl}$, the photon-ALP mixing probability $P_\mathrm{a\gamma}$ is suppressed by ${(\omega_\mathrm{pl}/m_\mathrm{a})}^{4}$. Close to resonance, i.e. as $m_\mathrm{a} \lesssim \omega_\mathrm{pl}$, $P_\mathrm{a\gamma}$ is enhanced \citep[see Sec. 4 of][]{Marsh17}. We pay special attention to resonant photon-ALP inter-conversion, conservatively excluding all resonant crossings (see Appendix~\ref{sec:appB_adaptiveTreatmentForResonances} for further discussion). 

\begin{figure}
    \centering
    \includegraphics[width=0.45\textwidth]{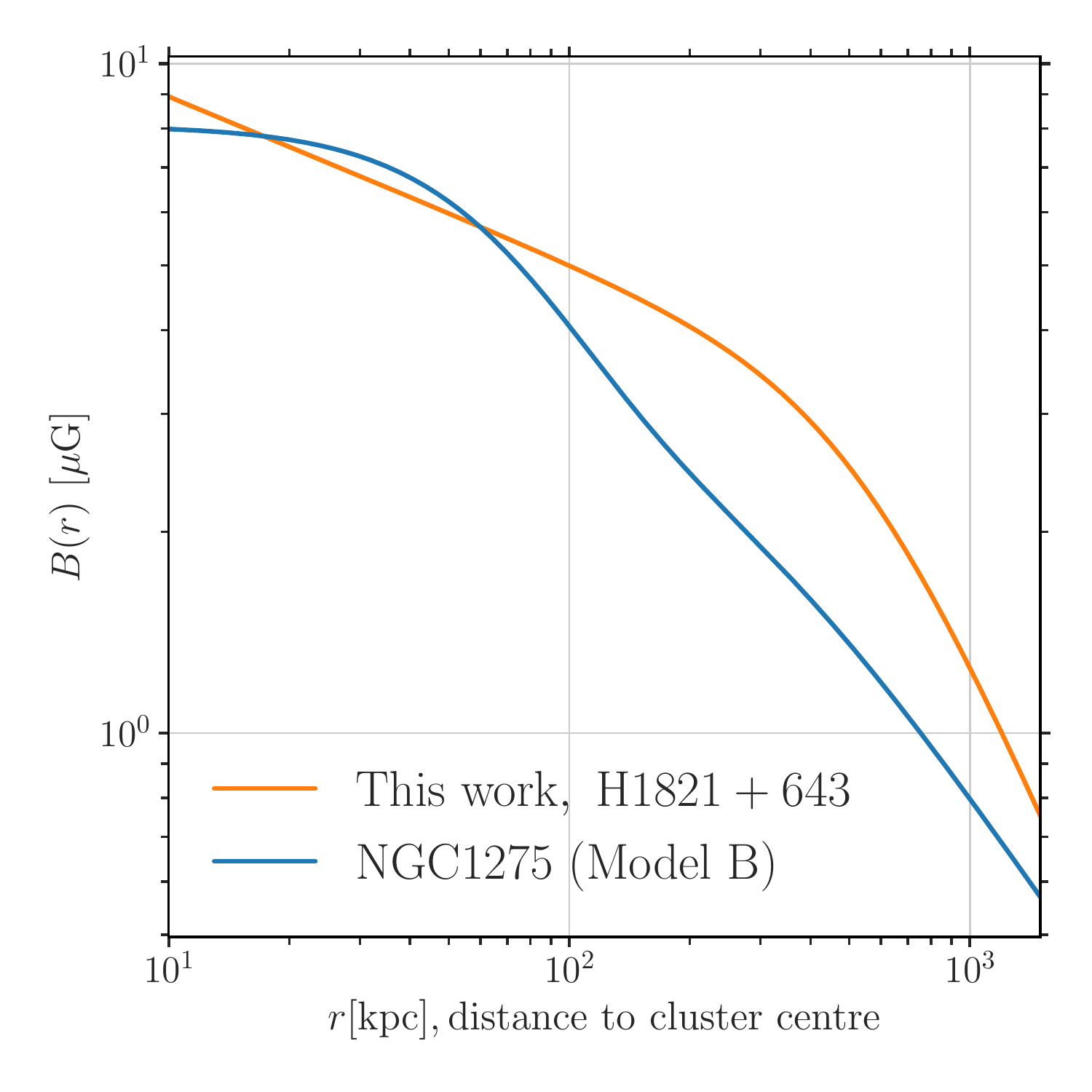}
    \caption{Estimated magnetic field strength across the sampled cluster volume, hosting the quasar H1821+643, in orange. For reference, we show the ICM field strength within the Perseus cluster predicted by Model B of \citet[][]{Reynolds20}, in blue, which hosts the NCG1275 central engine. We note that the relative errors on $B(r)$, derived from error propagation using the covariance matrix of our best fit to the \textsc{DSDEPROJ} data from \citet[][]{Russell10}, are below $0.2 \%$.}
    \label{figure:fig_myBr.pdf}
\end{figure}

Our cluster magnetic field modelling is motivated by Model B of \citet[][]{Reynolds20} and by \citet[][]{Marsh17}. We use a cell-based approach to generate 500 realisations of the cluster magnetic field by assuming that the thermal-to-magnetic pressure ratio $\beta$ remains constant within $10 \ \mathrm{kpc} - 1.5 \ \mathrm{Mpc}$ to the cluster centre. We assume a fiducial value of $\beta = 100$; see Sec. \ref{sec:sub5.2_BfieldModelAndLimitations} for further discussions. This yields a magnetic field intensity profile, $|{\textbf{B}(r)}|$, given by: \begin{equation}
\centering
\label{eq:BFieldIntensityAtClusterCenter}
     {|\textbf{B}(r)|}^{2} = P_\mathrm{th}(r) \times 4\pi / \beta
\end{equation} in \textit{cgs} units, where $P_\mathrm{th}(r)$ results from evaluating our best fit to the cluster thermal pressure at co-moving distance $r$ to the cluster centre, using Eq. \ref{eq:h1821_PRESSUREprofileFits}. The estimated field intensity within CL1821+643 is shown by Figure \ref{figure:fig_myBr.pdf}, which additionally shows the predicted field intensity profile within Perseus inferred by Model B of \citet[][]{Reynolds20}. We note that our field profile dominates within $r \in [1.0, 1.1) \times 10 \ \rm{kpc}$, and that its overall shape falls less steeply up to $1.5 \ \rm{Mpc}$ when compared to Model B of \citet[][]{Reynolds20}.

For each of the generated field configurations (which we refer to as \textit{Seeds}), we split the distance travelled by quanta into cells of thickness $ \Delta r \in [3.5, 10.0] \ \mathrm{kpc}$. For a given \textit{Seed}, the thickness of each of these cells, $\Delta r$, sets the scale over which the ICM magnetic field ($\textbf{B}$) is coherent. The sampling of coherence lengths is assigned following an underlying distribution function $p({\Delta r})\sim {\Delta r}^{-1.2}$. The direction of $\textbf{B}$ is set randomly and independently in each cell, so that \textbf{B} is able to have non-zero components along both directions perpendicular to the cluster line-of-sight. We take the ICM plasma frequency within each cell to be constant, and approximate it by evaluating Eq. \ref{eq:PlasmaFrequency} at $\Delta r/2$, which roughly corresponds to the average ICM plasma frequency within each cell. 

Our approach for describing the turbulent modes of the cluster field is approximate, since cell-based models are known to reproduce some, but \textit{not} all properties of turbulent fields \citep[refer to VII.B. of][]{Marsh2021_Fourier}. Nevertheless, it is aligned with the approach of other authors who embed a more detailed description of the stochastic nature of cluster fields \citep[e.g.][]{Angus_MagneticFields_ALPs}, and yet is relatively computationally inexpensive. We refer the reader to \citet[][]{Marsh2021_Fourier} for a discussion on the impact of different field modelling assumptions in environments suited to constrain ALPs.

We note that the magnetic field coherence lengths we consider are identical to those considered for Perseus in Model A of \citet[][]{Reynolds20} and in \citet[][]{Berg:2016_NGC1975}. In both studies, given the lack of detailed knowledge of the field structure within Perseus, coherence lengths were drawn based on the underlying probability distribution $p({\Delta r})$ of coherence lengths found for the Abell 2199 (A2199) cool-core cluster, following the findings of \citet[][]{Vacca12_Abell2199_PowerSpec}, who performed the highest quality Rotation Measure (RM) study that has been performed to date on a cool-core cluster. The authors found $p({\Delta r})$ from the best-fit 3-dimensional magnetic field power-law spectrum to RM observations of A2199, based on a fairly conservative field scaling with the thermal gas density. Furthermore, the field coherence length range considered by Model A of \citet[][]{Reynolds20} and \citet[][]{Berg:2016_NGC1975}, $\in [3.5, 10.0] \ \mathrm{kpc}$, was found by applying a cluster mass scaling term to the lower bounds of the minimum and maximum scales of the field fluctuations found in A2199 \citep[][]{Vacca12_Abell2199_PowerSpec}.

Theoretically, the coherence length, $\ell_\mathrm{c}$, of a cluster magnetic field should be related to the power spectrum of the turbulence in the ICM. A reasonable choice for the outer scale of the Kolmogorov contribution to the power spectrum in these stratified turbulent systems is the Ozmidov length scale $\ell_\mathrm{Oz}$, below which nonlinear advection dominates over the buoyant response of the ICM \citep[e.g. see discussions in][]{Zhuravleva2014_turbulentPerseusVirgo}. Using our best fits to the thermal profiles of CL1821+643 (see Sec. \ref{sec:s2.3_Russell10_descriptionAndTreatment}), assuming that the local turbulent dissipation rate is comparable to the local cooling rate, we estimate $\ell_\mathrm{Oz} \sim 15-50 \ \mathrm{kpc}$. For broadband Kolmogorov turbulence, $\ell_\mathrm{c}$ tends to $1/5$ of the maximum length scale of the turbulent power spectrum \cite[e.g.][]{harari2002}, so this Ozmidov length range would approximately correspond to a coherence length range of $\ell_\mathrm{c} \sim \ell_{\rm Oz}/5 \sim 3-10 \ \mathrm{kpc}$. This further justifies the range of cell sizes $\Delta r$ we have adopted for CL1821+643.

For each \textit{Seed} realisation, we generate a library of photon survival curves ($P_\mathrm{\gamma \gamma} \equiv 1 - P_\mathrm{a\gamma}$), assuming that as quanta emitted at the core of H1821+643 travel within $10 \ \mathrm{kpc} - 1.5 \ \mathrm{Mpc}$ to the cluster centre, they can convert into an ALP characterised by parameters $m_\mathrm{a}, g_\mathrm{a\gamma}$, and vice-versa. In our notation, $m_\mathrm{a}$ denotes the ALP rest mass, and $g_\mathrm{a\gamma}$, the ALP/two-photon coupling vertex. 

For a given \textit{Seed}, we compute the photon-ALP mixing probability for an initially unpolarised beam, for a grid of ALP models, characterised by $m_\mathrm{a}$, $g_\mathrm{a\gamma}$. We consider quanta energies $E= 1-100 \ \mathrm{keV}$, with $2 \ \mathrm{eV}$ energy resolution. We solve the Schrodinger-like equation by computing the transfer matrix in each cell, under the assumption that the magnetic field and plasma frequency are constant in each cell. The output state vector (amplitude and phase) from cell $j$ is then used as input to the calculation in cell $j+1$, where the subsequent cell has a different magnetic field strength, plasma frequency, and associated transfer matrix. This process continues until the photon-ALP beam is propagated to the virial radius. The magnetic field orientation is assigned randomly with the strength set from the field intensity profile at a given distance from the cluster centre (see Fig. \ref{figure:fig_myBr.pdf}). The details of our code will be described in a future paper (Matthews et al. in prep.), but our general procedure broadly follows other approaches described in the literature \citep[e.g.][]{DeAngelis:2011id,Marsh17,meyer2021,davies2021}. 

Our ALP survival curves omit the subleading effect of photon-axion inter-conversion across the large-scale Galactic magnetic field (GMF). We assess the impact of this omission using the GMF model of \citet[][]{JanssonFarrar2012} together with the photon-ALP mixing propagation code of \citet[][]{Meyer2014_GRF}. The maximum Galactic field strength in the direction of H1821+643 inferred is $\sim 0.8 \ \mu\mathrm{G}$ and rapidly falls off with distance from Earth. We find that the overall impact of the GMF is minor compared to the influence of the magnetised cluster and would only introduce a small phase shift in energy, roughly equivalent to including an extra $\sim 10 \ \mathrm{kpc}$ magnetised cell in the cluster outskirts.

The ALP models we generate span the parameter space $\mathrm{log}(m_\mathrm{a}/\mathrm{eV}) \in [-13.7, -10.1]$, $\mathrm{log}(g_\mathrm{a\gamma}/{\mathrm{GeV}}^{-1}) \in [-13.0, -10.2]$. Both $\mathrm{log}(m_\mathrm{a}/\mathrm{eV})$ and $\mathrm{log}(g_\mathrm{a\gamma}/{\mathrm{GeV}}^{-1})$ were sampled at $0.1 \ \mathrm{dex}$. This results in 1,073 different photon-ALP mixing models for each of the 500 realistic cluster field configurations considered. The lower bounds corresponding to each sampled variable reproduce photon-ALP oscillations at vanishing amplitude, and are therefore equivalent to photon-ALP mixing curves with $m_\mathrm{a} = 0$ and $g_\mathrm{a\gamma} = 0$, respectively.

We note that in Sec. \ref{sec:s5_Results_Discussion}, we present our inferred constraints on light ALPs within $\mathrm{log}(m_\mathrm{a}/\mathrm{eV}) \in [-30.0, -10.1]$, $\mathrm{log}(g_\mathrm{a\gamma}/{\mathrm{GeV}}^{-1}) \in [-19.0, -10.2]$. The lower mass bound corresponds to ALPs whose wavelength is equivalent to the size of the observable Universe, at which light ALPs would cease to qualify as DM candidates, and would instead qualify as DE fields. Moreover, $\mathrm{log}(g_\mathrm{a\gamma}/{\mathrm{GeV}}^{-1}) < -19.0$ embeds the phenomenology of ALPs at scales above the Planck energy, which would require the inclusion of the quantum effects of gravity for an adequate treatment. Furthermore, as described in the next section, to compute the posterior probabilities of all ALP models of $\mathrm{log}(m_\mathrm{a}/\mathrm{eV}) < -13.7$ and/or $\mathrm{log}(g_\mathrm{a\gamma}/{\mathrm{GeV}}^{-1}) < -13.0$, we take those inferred for the $\mathrm{log}(m_\mathrm{a}/\mathrm{eV}) = -13.7$ and/or $\mathrm{log}(g_\mathrm{a\gamma}/{\mathrm{GeV}}^{-1}) = -13.0$ model(s) as a proxy/proxies.

Figure \ref{figure:MixingCurves} shows the average survival curves of photons mixing with ALPs of $\mathrm{log}(m_\mathrm{a}/\mathrm{eV}) = -13.0$, $\mathrm{log}(g_\mathrm{a\gamma}/{\mathrm{GeV}}^{-1}) = -12.0$, having averaged over the 500 realistic cluster field configurations generated by our work (orange) and by Model B of \citet[][]{Reynolds20} (blue). We note the increase in the photon-ALP mixing strength predicted by our work within $1 - 2 \ \rm{keV}$, caused by the higher magnetic field intensity we predict within CL1821+643 at $r \in [1.0, 1.1) \times 10 \ \rm{kpc}$ to the cluster centre (see Fig. \ref{figure:fig_myBr.pdf}). However, for the ALP parameters illustrated in Fig. \ref{figure:MixingCurves}, photon-ALP mixing becomes more prevalent for photons emitted by NCG1275 as they travel through Perseus at energies $2 - 10 \ \mathrm{keV}$. This is expected given that Model B of \citet[][]{Reynolds20} accounted for the expected increase of field coherence lengths with distance from the cluster centre \citep[see Sec. 3 of][]{Reynolds20}. For further reference, Fig. \ref{figure:MixingCurves} depicts the expected survival curves for photons mixing with ALPs of $\mathrm{log}(m_\mathrm{a}/\mathrm{eV}) = -13.0$, $\mathrm{log}(g_\mathrm{a\gamma}/{\mathrm{GeV}}^{-1}) = -12.0$ for two random realisations of the field within CL1821+643 we have generated (solid and dash-dotted black lines). Clearly, regardless of the fact that both field realisations embed field geometries that reproduce the predicted field intensity profile within CL1821+643 (orange line in Fig. \ref{figure:fig_myBr.pdf}), their survival curves have a considerably different appearance. This highlights the importance of having generated 500 realistic realisations of the cluster field, as well as that of eliminating the dependence on the particular field configuration when computing posteriors on ALPs as follows. 

\section{Bayesian technique to infer posteriors}
\label{sec:s4_toInferPosteriors}
We now outline our route to producing the ALP constraint plot presented in Sec. \ref{sec:s5_Results_Discussion}. As described above, we start from our grid of photon survival curves for photons mixing with ALPs across: $\mathrm{log}(m_\mathrm{a}/\mathrm{eV}) \in [-13.7, -10.1]$, $\mathrm{log}(g_\mathrm{a\gamma}/{\mathrm{GeV}}^{-1}) \in [-13.0, -10.2]$. For each generated photon-ALP mixing curve featuring a specific ALP model -say, \textsc{J}-, we translate \textsc{J} into a multiplicative table model by using \textsc{xspec}'s \textit{flx2tab} command. We do this for all \textsc{J}s contained within each of the 500 generated cluster field configurations, and fit the model \textsc{CONST*TBABS*(PO[A]+DISKLINE+ZGAU+PO[B])*ALP(J)} to the HETG/LETG quasar spectrum. The \textit{C}-stats of all ALP models within a given \textit{Seed} are stored. Each ALP model has an associated \textit{redshift} free parameter, frozen at that of the quasar, $z = 0.299$.

The next step is to convert the \textit{C}-stat associated with each fitted ALP model, denoted by ALP parameters $(m_\mathrm{a}, g_\mathrm{a\gamma})$ and by a particular \textit{Seed} configuration $B_\mathrm{i}$, where $i \in [1,500]$, to a posterior probability in ALP parameter space. 

We employ the Bayesian procedure discussed in A2 of \citet[][]{Marsh17}. For a particular ALP and \textit{Seed} model (i.e. of fixed $m_\mathrm{a}$, $g_\mathrm{a\gamma}$; and $B_\mathrm{i}$, respectively) of an associated \textit{C}-stat $C(m_\mathrm{a}, g_\mathrm{a\gamma}, B_\mathrm{i})$, its associated posterior probability $p_\mathrm{post} (m_\mathrm{a}, g_\mathrm{a\gamma}, B_\mathrm{i})$ is given by: \begin{equation}
    \centering
    \label{eq: Posterior_probability_forGivenC}
    {p}_\mathrm{post}(m_\mathrm{a}, g_\mathrm{a\gamma}, B_\mathrm{i}) \propto \mathrm{exp} \biggl\{ \frac{ -(C(m_\mathrm{a}, g_\mathrm{a\gamma}, B_\mathrm{i}) - C_\mathrm{fid})}{2} \biggl\},
\end{equation} where $C_\mathrm{fid} = 5490.22$ is the $C$-stat value associated with our best-fit fiducial model for the combined quasar spectrum (see Sec. \ref{sec:s2.2_FiducialQuasarMod_h1821+643}). Hence, the fitted ALP models for which $C(m_\mathrm{a}, g_\mathrm{a\gamma}, B_\mathrm{i}) < C_\mathrm{fid}$ will be the most relevant in populating the underlying probability distribution of ALPs. Eq. \ref{eq: Posterior_probability_forGivenC} also hints that subtle changes in $C$-stats are likely to propagate significantly across the distribution of ALP posteriors. 

We normalise the ALP posterior distribution across $\mathrm{log}(m_\mathrm{a}/\mathrm{eV}) \in [-30.0, -10.1]$, $\mathrm{log}(g_\mathrm{a\gamma}/{\mathrm{GeV}}^{-1}) \in [-19.0, -10.2]$ by assuming flat priors, using:\begin{equation}
    \centering
    \label{eq: Posterior_probability_Normalized} 
    \sum_{\mathrm{log}(m_\mathrm{a}/\mathrm{eV})} \ \sum_{\mathrm{log}(g_\mathrm{a\gamma}/\mathrm{GeV}^{-1})} \sum_{i = 1}^{500} {p}_\mathrm{post}(m_\mathrm{a}, g_\mathrm{a\gamma}, B_\mathrm{i}) = 1 ,
\end{equation} where we sum over all ALP mass and coupling values considered, as well as over the 500 $B_\mathrm{i}$ realisations of the cluster field we have generated. For a given ALP model quantified by variables $(m_\mathrm{a}, g_\mathrm{a\gamma})$, we remove the dependence on the magnetic field model by marginalising over these configurations, following: \begin{equation}
    \centering
    \label{eq: Posterior_probability_forHavingMarginalizedOverSeeds}
    {p}_\mathrm{post}(m_\mathrm{a}, g_\mathrm{a\gamma}) \propto \sum_{i = 1}^{500} {p}_\mathrm{post}(m_\mathrm{a}, g_\mathrm{a\gamma}, B_\mathrm{i}) ,
\end{equation} where the summation is performed over all $i \in [1,500]$ realisations of the cluster field. We find the proportionality constants appearing in Eqs. \ref{eq: Posterior_probability_forGivenC} and \ref{eq: Posterior_probability_forHavingMarginalizedOverSeeds} by using Eq. \ref{eq: Posterior_probability_Normalized}. 

Armed with our marginalised posteriors $p_\mathrm{post}(m_\mathrm{a}, g_\mathrm{a\gamma})$ across a grid of ALP parameters, we can now form confidence regions at a given level of significance. In order to constrain ALPs, our aim is to find ALP models that concentrate the highest probability across the underlying parameter space. For this reason, we first sort all $p_\mathrm{post}(m_\mathrm{a}, g_\mathrm{a\gamma})$ by decreasing values of the posterior. The only remaining ambiguity when computing cumulative probabilities is how to handle ALP models that have identical posteriors. We assign the mean cumulative to all such models, although equivalently, one could sort all ALP grid models by increasing values of $m_\mathrm{a}$ and $g_\mathrm{a\gamma}$. We note that this choice does not affect our results. We proceed by finding all ALP models comprised within the $2\sigma$ and $3\sigma$ CLs, of associated cumulative posteriors $j = 95.5\%, \ 99.7\%$, respectively, by summing over all sorted $p_\mathrm{post}(m_\mathrm{a}, g_\mathrm{a\gamma})$ up to $j$.

\section{Results and discussion}
\label{sec:s5_Results_Discussion}

\subsection{Results and reflection on subsequent cluster ALP studies}
\label{sec:sub5.1_ResultsAndReflectionOnFutureStudies}
\begin{figure*}

 \center

  \includegraphics[width=.95\textwidth]{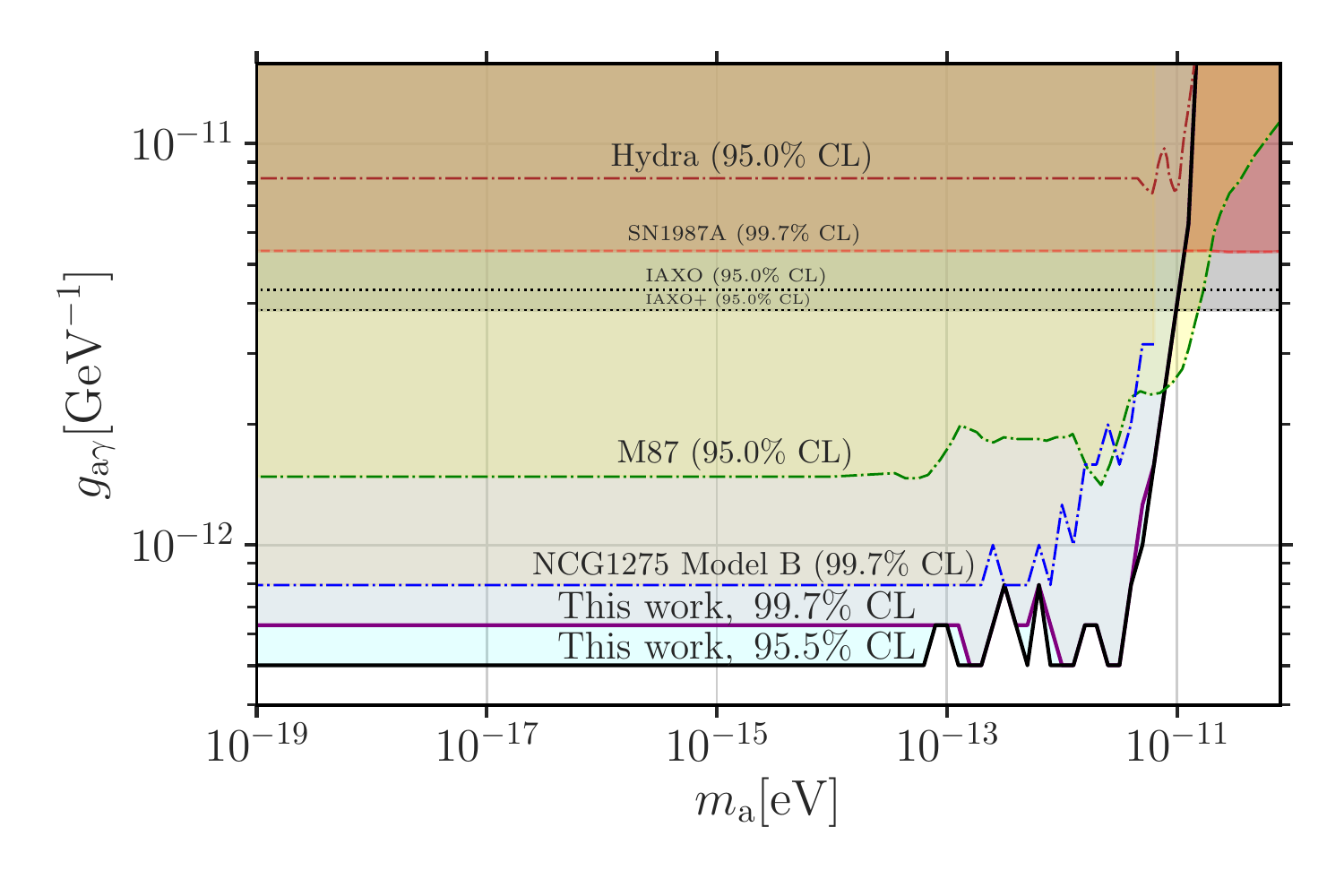}
  \caption{Exclusion curve we infer when marginalising over the 500 generated realisations of the field within CL1821+643 (thick black and magenta lines at $95.5\%$, $99.7\%$ credibility, respectively). These constraints assume a thermal-to-magnetic pressure ratio of $\beta=100$. For a different choice of $\beta$, say $\beta^\prime$, our exclusion curves are scaled in $g_\mathrm{a\gamma}$ by a factor of $\sqrt{\beta^\prime/100}$ (refer to Sec. \ref{subsubsection:assumption_on_beta}). The shaded regions above all curves underlie the parameter space excluded by any of the studies shown, at the credibility region (CL) quoted above each exclusion curve. Our results improve from the upper bound on $g_\mathrm{a\gamma}$ set by the previous most sensitive studies based on different AGN-cluster systems (dash-dotted lines, refer to Tab. \ref{table:Tab1_Constraints_centralAGN_CoolCore}). We show the projected sensitivity of the next-generation axion helioscope IAXO+, and its upgrade IAXO (dotted lines). We note that the credibility region attributed to SN1987A was extrapolated from $m_\mathrm{a} \lesssim 4.4 \times 10^{-10} \ \mathrm{eV}$, from where ALP-cluster studies are unable to provide constraints, since photon-ALP mixing is suppressed. The $99.7\%$ CL quoted for SN1987A corresponds to the upper limit on the fluence during the neutrino burst.}
\label{figure:Results_ConfidenceLimits_ALPexclusion}
\end{figure*}

Our fiducial constraints on the ALP-photon coupling constant assuming $\beta=100$ are shown in Fig. \ref{figure:Results_ConfidenceLimits_ALPexclusion}. At $99.7\%$ confidence, we set a strong upper bound on the coupling of ALPs to the electromagnetic force, namely $g_\mathrm{a\gamma} > 6.3 \times 10^{-13} \ {\mathrm{GeV}}^{-1}$, for most light ALPs of masses $m_\mathrm{a} < 10^{-12} \ \mathrm{eV}$. At $95.5\%$ credibility, we exclude $g_\mathrm{a\gamma} > 5.0 \times 10^{-13} \ {\mathrm{GeV}}^{-1}$ within the same mass range.

Our constraints exceed the projected sensitivity of the next-generation axion helioscope, the International AXion Observatory IAXO, for these very light ALPs, as well as that of its upgrade IAXO+ \citep[see Fig. 14 of][]{IAXO19_ProjectedSensitivity}, by almost an order of magnitude. Additionally, Fig. \ref{figure:Results_ConfidenceLimits_ALPexclusion} shows the upper bound on $g_\mathrm{a\gamma}$ inferred from the absence of a gamma-ray signal attributed to ALPs from SN1987A \citep[][]{SN1987Bounds}. 

The limits obtained in our work are a modest improvement over the best previous result based upon \textit{Chandra} observations of the Perseus cluster \citep[][]{Reynolds20}, and constitute the tightest constraints to date on light ALPs of $\mathrm{log}(m_\mathrm{a}/\mathrm{eV}) < -12.0$, albeit equally relying on the assumption that $\beta = 100$ up to the cluster virial radius. 

The suitability of a particular AGN-cluster system for ALP studies depends on the $\mathrm{S/N}$ photon statistics; the effect of pile-up, which can harden the AGN spectrum; and the disentanglement between cluster and intrinsic AGN emission. Unfortunately, even if these effects are mitigated, cluster ALP searches are, and will continue to be strongly limited by the energy-dependent calibration of X-ray telescopes. 

\subsection{Magnetic field modelling}
\label{sec:sub5.2_BfieldModelAndLimitations}
Our study is based on similar cluster field modelling considerations as Model B of \citet[][]{Reynolds20}. Despite the fact that our study is dominated by different systematics, we find an upper bound on $g_\mathrm{a\gamma}$ consistent with that inferred by Model B of \citet[][]{Reynolds20}. The latter excluded $g_\mathrm{a\gamma} > 8 \times 10^{-13} \ {\mathrm{GeV}}^{-1}$ ($99.7\%$ CL) based on a deep \textit{Chandra} observation of its central engine, NGC1275. 

The most direct probes of the magnetic field strength in the ICM are Faraday Rotation Measures (RMs) of embedded or background radio sources. Even though there are no RM measurements available for CL1821+643, it is still useful to make predictions of the cluster RMs using our fiducial field model. 
The distribution of RMs from 500 generated field realisations for CL1821+643 is approximately Gaussian, with a mean close to zero, and a standard deviation of $11\,232 \ \mathrm{rad\,m^{-2}}$.
The median absolute RM of this set of realisations is $7\,471 \ \mathrm{rad\,m^{-2}}$. Therefore, the absolute RM of CL1821+643 is potentially quite large, which would require high frequency-resolution radio polarimetry to uncover.

Broadly, our approach to modelling the cluster magnetic field is conservative as follows. Firstly, our photon-ALP mixing grid accounts for AGN photons which travel within $10 \ \mathrm{kpc} - 1.5 \ \mathrm{Mpc}$ from the cluster centre. In general, one would expect the photon-ALP mixing probability to be enhanced within the inner $10 \ \mathrm{kpc}$ of the cluster, given the large field intensities. Secondly, we have implemented a treatment to identify and suppress ALP-photon resonant inter-conversion for all generated photon-ALP mixing models (refer to Appendix~ \ref{sec:appB_adaptiveTreatmentForResonances}). Finally, we note that in general, ICM field coherence lengths grow with distance from the cluster centre, and our coherence length sampling has \textit{not} accounted for this effect.

\subsubsection{Sensitivity to the value of $\beta$}
\label{subsubsection:assumption_on_beta}

In our work, we describe the turbulent field modes by splitting the cluster line-of-sight into a series of cells within which the field is constant. The radial dependence of the field intensity within a given cell is found by using our best fit to the cluster thermal pressure profile (Appendix~ \ref{sec:sec6.1_thermalPropertiesOfHostCluster}), under the assumption that $\beta = 100$ up to a maximum radius, $\sim 1.5 \ \mathrm{Mpc}$. 

As discussed by, e.g., \citet[][]{Donnert2018_MagneticFieldReview}, $\beta \sim 100$ has been adopted as the fiducial value for the thermal-to-magnetic pressure ratio in clusters. While this is motivated by RM observations \citep[][]{Taylor06, Bohringer2016_obs_betaValue}, it also naturally arises from the assumption of equipartition between kinetic and magnetic energies in the ICM turbulence, coupled with observational estimates of the turbulent velocity fluctuations \citep[e.g.][]{Zhuravleva2014_turbulentPerseusVirgo}. However, given the imprecise nature of these results and the possibility of cluster-to-cluster variations, it is useful to discuss how our results change if we were to adopt a different value of $\beta$.

In the interaction Lagrangian (Eq. \ref{eq: Lagrangian_vertex}) that gives rise to ALP-photon mixing, the photon-ALP coupling and the component of the magnetic field aligned with the photon polarisation -- $B_i$ with $i \in (x,y)$ -- only appear together as the product $g_\mathrm{a\gamma}B_i$. Thus, for a given spatial field structure and for given cluster profiles, survival curves for different field strengths $|{{\textbf{B}}}|$ are identical if the photon-ALP coupling is scaled by $g_\mathrm{a\gamma}\propto 1/|{{\textbf{B}}}|\propto \sqrt{\beta}$. Therefore, the constraints shown by Figs. \ref{figure:Results_ConfidenceLimits_ALPexclusion} and \ref{figure:Results_LETG_vs_HETG} can be scaled to any value of the thermal-to-magnetic pressure ratio, say $\beta^\prime$, by multiplying $g_\mathrm{a\gamma}$ by $\sqrt{\beta^\prime/100}$.

\subsubsection{Synchrotron emissivity and minimum energy argument}
\label{subsubsection:MinEnergyArgument}

While there are no existing RM-based estimates of the magnetic field within CL1821+643, this cluster does contain two radio synchrotron emitting structures that can be used to obtain an independent estimate of the cluster field.

On large scales, CL1821+643 hosts a large radio halo. Viewed at 323\,MHz, this extends across a projected region of 1.1\,Mpc centered on the cluster core with a total flux density of $F_{323\,{\rm MHz}} = 62 \ \pm \ 4 \ \mathrm{mJy}$ \citep[][]{CL1821+643_Bonafede14}. \citet[][]{Kale16_CL1821+643} present $610 \ \mathrm{MHz}$ observations of this same structure, finding a spectral index of $\alpha \approx 1$ and a roughly elliptical source with major and minor axes of $890 \ \mathrm{kpc}$ and $450 \ \mathrm{kpc}$, respectively. We estimate the magnetic field within this structure from standard synchrotron theory \citep[refer to Sec. \ref{sec:app_c_equipartition_derivation}, see also][]{RybickiAndLightman}, making the assumptions that the radio halo (i) is uniform and has unity filling factor throughout an ellipsoidal form with axes 890\,kpc $\times$ 450\,kpc $\times$ 670\,kpc (where the third axis is the arithmetic mean of the other two), (ii) has equal energy density in relativistic protons and electrons ($k_{\rm prot}=1$), and (iii) minimises the sum of the energy densities of the relativistic particles and magnetic field at the observed synchrotron emissivity. Taking the electron energy spectrum to have an index $p=3.14$ \citep[the arithmetic mean of the lower and upper values of $p$ found by][]{CL1821+643_Bonafede14}, extending down to a minimum Lorentz factor of $\gamma_{\rm min}=1$, we find a field estimate of $0.3 \ \mathrm{\mu G}$, in broad agreement with \cite{Kale16_CL1821+643}. 

This magnetic field estimate is conservative, however, and the actual field strength associated with the observed radio emission may be significantly higher, as follows. Firstly, there may be significantly more energy in relativistic protons than relativistic electrons, a possibility highlighted by the fact that the electron energy spectrum is steep and hence likely influenced by radiative losses. Plausible values of the relativistic proton-to-electron energy density are $k_{\rm prot}\sim 50$, or even higher \citep[e.g. see Fig. 3 of ][]{CrostonHardcastleKprot}. Adopting $k_{\rm prot} \in [50, 100)$ and noting that the inferred magnetic field is $\propto (1+k_{\rm prot})^{2/7}$ (see Eq. \ref{eq:Beq_fromRybickiandLightman}), this raises our field estimate to $1\,\mu\rm G$. Secondly, the assumption of unity filling factor throughout the ellipsoid is likely to be invalid. The size of this radio halo is comparable to the radio relics seen in the outskirts of some clusters which are associated with cluster merger shocks and which possess highly elongated forms, possible suggesting a quasi-planar-like geometry. If the giant radio halo in CL1821+643 is indeed a radio relic viewed face-on, the effective filling factor $f$ could be an order of magnitude smaller than unity. At 500\, $\rm kpc$, this would further increase the inferred field strength to $2 \ \mu \rm G$, which roughly corresponds to the field strength predicted by our fiducial field model at $500 \ \rm kpc$ to the cluster centre.

Additionally, in the central regions of CL1821+643, there is a FR-I central radio structure that is likely associated with the quasar. \citet[][]{BlundellandRawlings_2001_FRI_inCL1821+643} report a total flux density of $F_{5 \ \mathrm{GHz}} = 44.4 \ \rm mJy$. Repeating the argument above, treating the FR-I source as a sphere of radius of $140 \ \mathrm{kpc}$, we derive a conservative ($k_{\rm prot}=1$; unity filling factor) magnetic field estimate of $1 \ \mathrm{\mu G}$. Furthermore, additional pressure from relativistic protons and deviations from the assumed geometry could readily increase the estimated field at 140\,kpc to $3-5\,\mu$G.

\subsection{Sensitivity to inner and outer radial cutoffs}
\label{sec:sub5.3_SensitivityToRadialCutoffs}

Our results are marginally sensitive to the minimum (maximum) radius from (up to) which we compute photon-ALP mixing. As introduced in Sec. \ref{sec:s3_ALPSurvivalCurves}, throughout our study, we have computed inter-conversion probabilities $P_\mathrm{a\gamma}$ across a grid of ALP parameters $m_\mathrm{a}, g_\mathrm{a\gamma}$ within $10 \ \mathrm{kpc} - 1.5 \ \mathrm{Mpc}$ to the cluster centre. The former corresponds to the radial distance to the centre of Perseus from which Model A of \citet[][]{Reynolds20} computed $P_\mathrm{a\gamma}$; the latter is expected to be located within the lower bound of the estimate of the virial radius of CL1821+643 given in Sec. 5.3 of \cite{Russell10}. We note that the \textsc{DSDEPROJ} data from \citet[][]{Russell10} we have employed commences at $24 \pm 2 \ \mathrm{kpc}$ and terminates at $661 \pm 4 \ \mathrm{kpc}$ to the centre of CL1821+643 (see Fig. \ref{figure:VerticallyStackedFig_H1821_profiles} and refer to Sec. \ref{sec:s2.3_Russell10_descriptionAndTreatment}). To test the sensitivity to these adopted values, we recalculated $P_\mathrm{a\gamma}$ for a range of masses at $g_\mathrm{a\gamma} = 10^{-12} \ {\mathrm{GeV}}^{-1}$ and examined the mean inter-conversion probability (calculated over 500 field realisations) as a function of radius.

We find that setting the maximum radius to $661 \ \mathrm{kpc}$ results in an inter-conversion probability typically $20\%$ to $30\%$ lower for $m_\mathrm{a}<10^{-12}$\,eV. Given a specific set of ALP parameters and a given cluster field realisation, the change in path length travelled along the quasar line-of-sight effectively translates into a phase shift in $P_\mathrm{a\gamma}$. Averaging over field configurations, the discrepancy is smaller at higher $m_\mathrm{a}$. Since the mixing probability scales as $\sim g_\mathrm{a\gamma}^2$, decreasing the maximum radius to $661 \ \mathrm{kpc}$ might be expected to weaken the limits on $g_\mathrm{a\gamma}$ by $\sim10\%$ for $m_\mathrm{a} < 10^{-12} \ \mathrm{eV}$. Broadly, for all ALPs with $m_\mathrm{a} < \omega_\mathrm{pl}$, i.e. away from the resonant inter-conversion regime, the sensitivity to the maximum radius becomes slightly stronger at higher photon energies, where our data have less statistical power.

On the other hand, we find that, if averaging over field realisations, $P_\mathrm{a\gamma}$ is conserved when setting the minimum radius to, for example, both $1 \ \mathrm{kpc}$ and $20 \ \mathrm{kpc}$ for low mass ALPs ($m_\mathrm{a} < 10^{-12} \ \mathrm{eV}$). For $m_\mathrm{a} \sim 10^{-11} \ \mathrm{eV}$, predominantly at energies above the soft band, our constraints on $g_\mathrm{a\gamma}$ would be weakened by up to $10\%$ had we set the minimum radius to $20 \ \mathrm{kpc}$. Furthermore, we can test the sensitivity of $P_\mathrm{a\gamma}$ to the extrapolation in the inner regions of the cluster. To do this, we compared $P_\mathrm{a\gamma}$ calculated with a minimum radius $1 \ \mathrm{kpc}$ using two different extrapolations: one with the adopted $B(r)$ profile and another with the field within $22 \ \mathrm{kpc}$ uniformly set to $B(r = 22 \ \mathrm{kpc})$. In this case, $P_\mathrm{a\gamma}$ remains almost identical across the whole grid of ALP parameters we have sampled, since this procedure only slightly modifies $B(r)$ in a small region of the total path length travelled by AGN photons along the line-of-sight.

\subsection{Fiducial quasar model}
\label{sec:sub5.4_SensitivityToPriors}
\begin{figure*}
 \center
  \includegraphics[width=.95\textwidth]{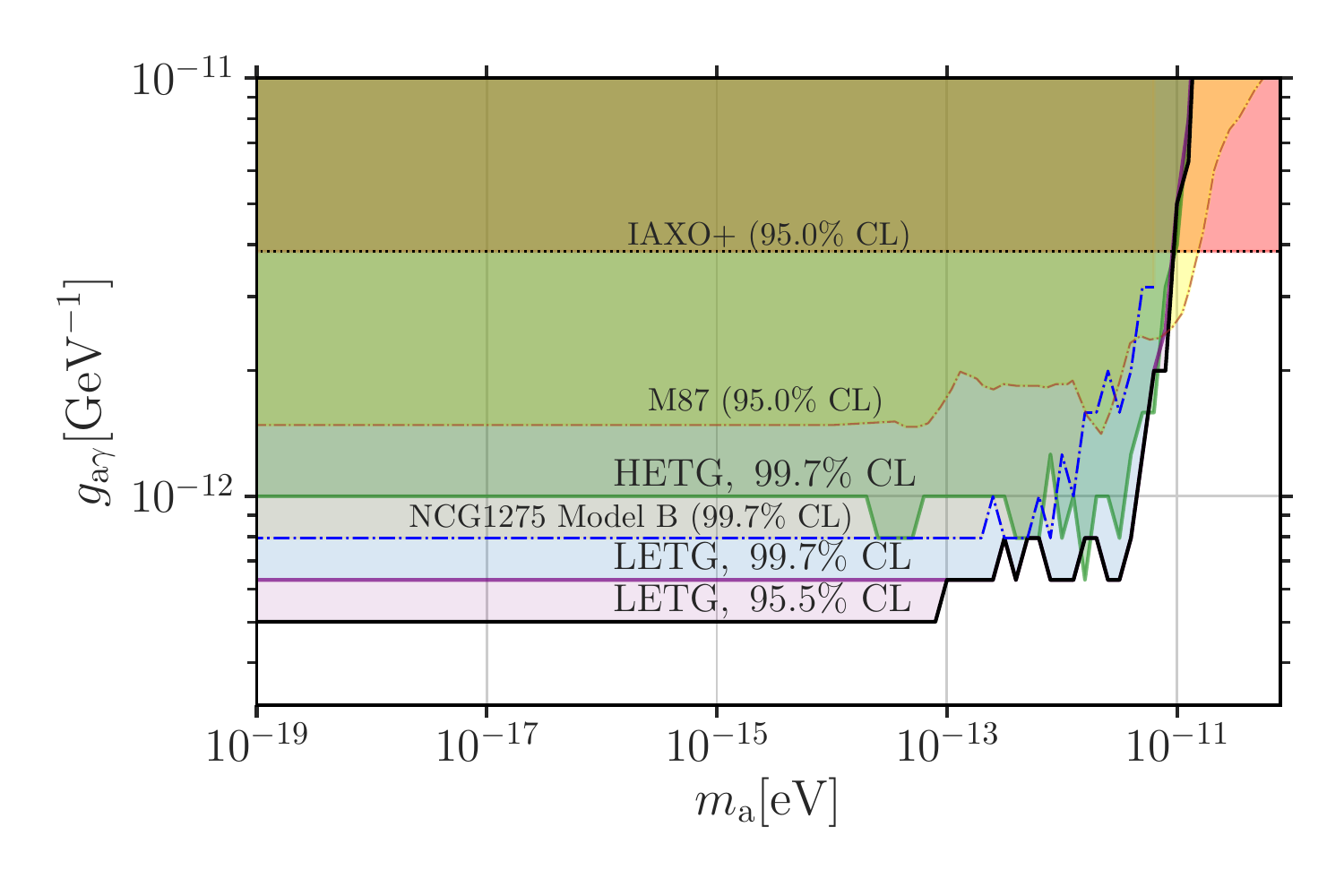}
  \caption{Exclusion curve we infer when fitting ALPs on the individual sets of HETG/LETG observations we have employed to extract our combined constraints shown in Fig. \ref{figure:Results_ConfidenceLimits_ALPexclusion}. The shaded regions above all curves underlie the parameter space excluded by any of the studies shown, at the credibility region (CL) quoted above each exclusion curve. We include the projected sensitivity of the upgrade of the next-generation axion helioscope IAXO, namely IAXO+ \citep[][]{IAXO19_ProjectedSensitivity}, as well as the results from Model B of \citet[][]{Reynolds20} and \citet[][]{Marsh17}. Clearly, our combined HETG/LETG analysis is driven by the set of LETG observations employed (which add up to a total exposure of $471.4 \ \rm{ks}$ over the $571 \ \rm{ks}$ total on-source exposure employed in our combined analysis). These constraints assume a thermal-to-magnetic pressure ratio of $\beta=100$. For a different choice of $\beta$, say $\beta^\prime$, our exclusion curves are scaled in $g_\mathrm{a\gamma}$ by a factor of $\sqrt{\beta^\prime/100}$ (refer to Sec. \ref{subsubsection:assumption_on_beta}).}
  \label{figure:Results_LETG_vs_HETG}
\end{figure*}

The upper bound on $g_\mathrm{a\gamma}$ inferred by fitting our fiducial quasar model to the individual HETG/LETG datasets employed throughout our analysis, following Tabs. \ref{table:Details_bestFitParValues} and \ref{table:FinalDetailsFiducialModel}, is shown by Fig. \ref{figure:Results_LETG_vs_HETG}. Interestingly, the upper bound inferred by the LETG data coincides with and therefore drives the results inferred with our combined HETG/LETG quasar analysis. This is unsurprising given that fitting our baseline quasar model to the LETG dataset results in spectral distortions $< 1.2 \%$ (see Fig. \ref{figure:myModFiducial}). On the other hand, the HETG-only runs exclude $\mathrm{log}(g_\mathrm{a\gamma}/{\mathrm{GeV}}^{-1}) > -12.0$ at $99.7\%$ CL for $\mathrm{log}(m_\mathrm{a}/\mathrm{eV}) < -12.0$.

We note that despite the constraints attributed to the HETG quasar observation being less sensitive to constraining ALPs than those inferred by the LETG, the inclusion of the former has been important for the reasons that follow. Firstly, the overall photon statistics were improved due to an increased on-source exposure. Secondly, the grating instruments mounted onto \textit{Chandra}'s HETG target higher energy ranges. Generally, one would expect the amplitude of the modulations induced by photon-ALP mixing to increase with energy, as shown by Fig. \ref{figure:MixingCurves}. Having incorporated the HETG data in our analysis has improved the upper bound we infer on $g_\mathrm{a\gamma}$ at masses $\mathrm{log}(m_\mathrm{a}/\mathrm{eV}) \in (-13.0, -12.0)$.

Our fiducial quasar model describes the combined HETG/LETG observation of H1821+643 through a double power law continuum. The reflection component is described with the use of the \textsc{diskline} and a \textsc{zgau} models. We account for potential quasar spectral variability by letting the intrinsic AGN and reflection model components vary amongst the two datasets, since they correspond to observations of the source taken at different time periods. When fitting ALPs, our baseline quasar model assumes that the inclination of the accretion disc is frozen at the best-fit value. This was motivated by the degeneracies resulting from fitting the double power law to the continuum. Given that we have a good description of the quasar spectrum, presenting spectral distortions only $< 2.5\%$, these degeneracies do \textit{not} affect our results.

\subsection{Sensitivity to priors and other assumptions}
\label{sec:sub_SensitivityToPriors}
We recall that with our analysis, we are able to constrain generic ALPs within $\mathrm{log}(m_\mathrm{a}/\mathrm{eV}) \in [-30.0, -10.1]$, $\mathrm{log}(g_\mathrm{a\gamma}/{\mathrm{GeV}}^{-1}) \in [-19.0, -10.2]$. Our results are almost insensitive to our choice of priors. Suppose, for instance, that our Bayesian technique for posterior inference (described in Sec. \ref{sec:s4_toInferPosteriors}) was employed only within the region of ALP parameter space for which we have generated photon-ALP mixing models, i.e. within $\mathrm{log}(m_\mathrm{a}/\mathrm{eV}) \in [-13.7, -10.1]$, $\mathrm{log}(g_\mathrm{a\gamma}/{\mathrm{GeV}}^{-1}) \in [-13.0, -10.2]$. In this scenario, our $99.7\%$ credibility region would have excluded all photon-ALP couplings $g_\mathrm{a\gamma} > 6.3 \times 10^{-13} \ {\mathrm{GeV}}^{-1}$ for most ALP masses $m_\mathrm{a} < 10^{-13} \ \mathrm{eV}$. At $2\sigma$ CL, the inferred constraints are identical to the $95.5\%$ credibility region shown by Fig. \ref{figure:Results_ConfidenceLimits_ALPexclusion}.

Finally, we note that the uncertainty associated with our fits to the thermal properties of CL1821+643 (Sec. \ref{sec:s2.3_Russell10_descriptionAndTreatment}) contribute to those associated with our inferred upper bound on $g_\mathrm{a\gamma}$ to sub-leading order.

\section{Conclusions}
\label{sec:s5_Conclusions}

In conclusion, our study excludes all ALP-photon couplings $g_\mathrm{a\gamma} > 6.3 \times 10^{-12} \ {\mathrm{GeV}}^{-1}$ for most light ALPs of masses $m_\mathrm{a} < 10^{-12} \ \mathrm{eV}$ ($99.7\% \ \mathrm{CL}$). This has slightly improved on the previous best result on such ALPs, based on a comparative magnetic field model. We have employed high-resolution \textit{Chandra} Transmission Spectroscopy of the powerful radio-quiet quasar H1821+643, located at the centre of the CL1821+643 massive cool-core cluster. Using a combined HETG/LETG deep observation of this source, of $571 \ \mathrm{ks}$ of exposure, we have extracted its highest quality existing spectrum free from photon pile-up, presenting spectral distortions below $2.5\%$. 

Our result illustrates that thermally rich clusters with X-ray bright central AGNs could be excellent targets of further ALP studies. However, future X-ray ALP studies performed with CCD detectors, even if schemed to circumvent spectral contamination of the AGN by cluster emission, as well as to minimise the effects of photon pile-up, will be strongly limited by knowledge on detector calibration.

Promisingly, \citet[][]{Conlon_2017_ProjectedBoundsOnAthena} simulated a $200 \ \mathrm{ks}$ \textit{Athena} observation of NGC1275 on the basis of a nominal Auxiliary Response File (ARF) \citep[][]{NominalCalibrationOfAthenaXRayTelescope}, to infer a projected upper bound $g_\mathrm{a\gamma} \lesssim 1.5 \ \times 10^{-13} \ {\mathrm{GeV}}^{-1}$ for ALPs of $m_\mathrm{a} \lesssim 10^{-12} \ \mathrm{eV}$ ($95\% \ \mathrm{CL}$). \textit{Athena}'s X-ray Microcalorimeter Spectrometer (XMS) will have a spectral resolution of $2.5 \ \mathrm{eV}$, which will enhance the discrimination against ALP models suited to describe the spectral distortions attributed to an AGN. However, a rigorous analysis of \textit{Athena}'s potential for these ALP searches must account for realistic calibration uncertainties in the mirror effective area and detector response. This will be the subject of a future publication. 

\section*{Acknowledgements}
We thank the anonymous reviewer for a constructive and helpful report. J.S.R acknowledges the support from the Science and Technology Facilities Council (STFC) under grant ST/V50659X/1 (project reference 2442592). J.H.M acknowledges a Herchel Smith Fellowship at Cambridge. C.S.R. thanks the STFC for support under the Consolidated Grant ST/S000623/1, as well as the European Research Council (ERC) for support under the European Union’s Horizon 2020 research and innovation programme (grant 834203). H.R.R acknowledges the support from an STFC Rutherford Fellowship and an Anne McLaren Fellowship. R. N. S. acknowledges support from NASA under the Chandra Guest Observer Program (grants G08-19088X and G09-20119X). D.M. is supported by the European Research Council under Grant No. 742104 and by the Swedish Research Council (VR) under grants 2018-03641 and 2019-02337. We would like to thank Pierluca Carenza, Jamie Davies and Andy Fabian for helpful discussions. This work was performed using resources provided by the Cambridge Service for Data Driven Discovery (CSD3) operated by the University of Cambridge Research Computing Service (\url{www.csd3.cam.ac.uk}), provided by Dell EMC and Intel using Tier-2 funding from the Engineering and Physical Sciences Research Council (capital grant EP/P020259/1), and DiRAC funding from the Science and Technology Facilities Council (\url{www.dirac.ac.uk}).

\section*{Data Availability}
The raw X-ray data on which this study is based is available in the public data archives of the Chandra Science Center. The reduced data products used in this work may be shared on reasonable
request to the authors.

\bibliographystyle{mnras}
\bibliography{myReferences}

\appendix
\section{Fits to thermal properties of the host cluster}
\label{sec:sec6.1_thermalPropertiesOfHostCluster}

\begin{figure}
    \centering
    \includegraphics[width=0.45\textwidth]{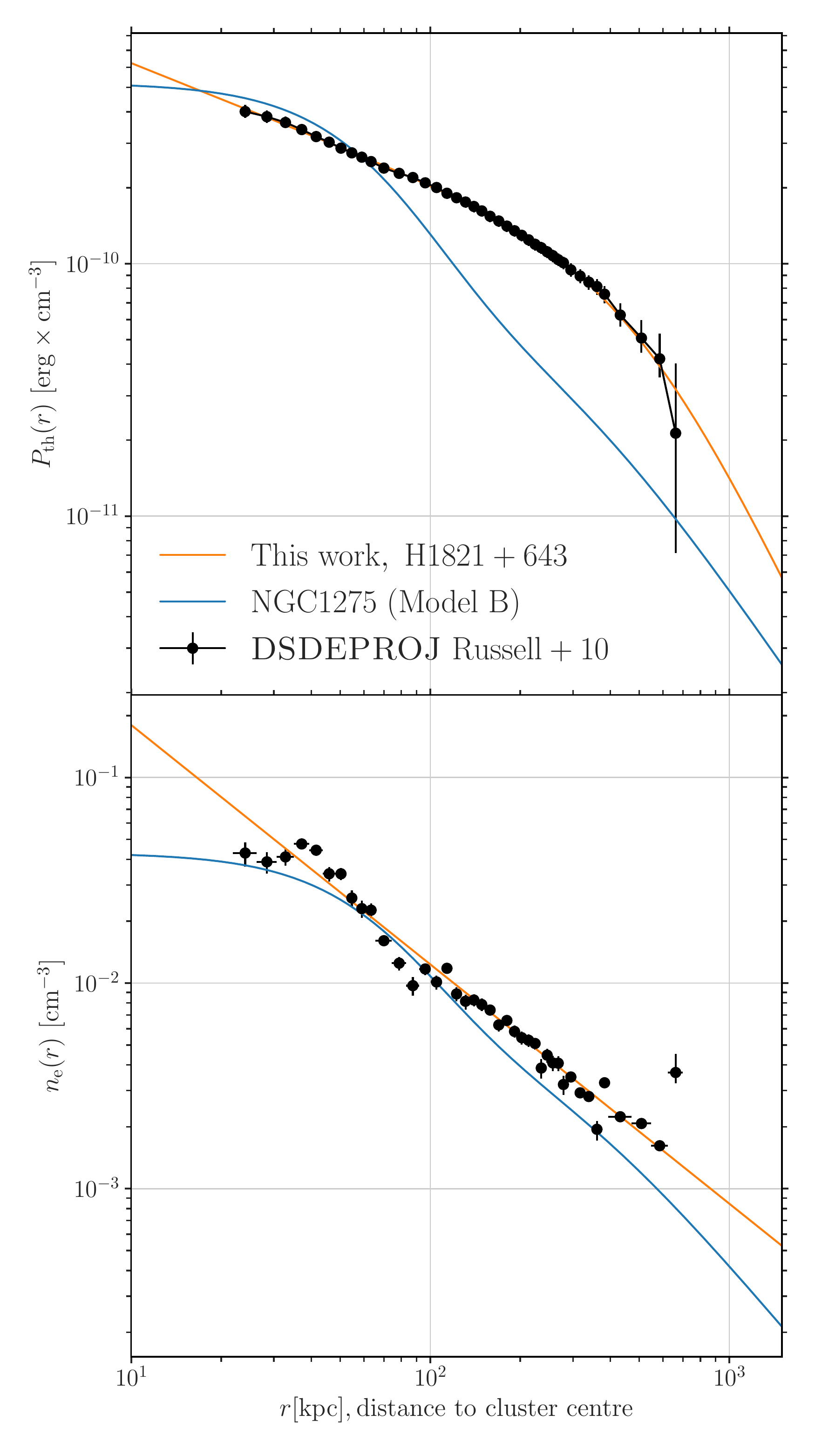}
    \caption{Our best fits (in orange) to the \textsc{DSDEPROJ} \citet[][]{Russell10} data (black dots) of the host cluster: $P_\mathrm{th}(r)$ (thermal pressure profile; upper panel) and $n_\mathrm{e}(r)$ (electron number density profile; lower panel). We also show the profiles assumed by Model B of \citet[][]{Reynolds20}, where that in the lower panel corresponds to the analytic density profile inferred by \citet[][]{Churazov+03} from fitting Perseus data. The increase in the outermost radial bin present in the \textsc{DSDEPROJ} data is an artifact of the deprojection process. Our best fit parameter values are indicated in Sec. \ref{sec:s2.3_Russell10_descriptionAndTreatment}.}
    \label{figure:VerticallyStackedFig_H1821_profiles}
\end{figure}

We introduce our best fits to the \textsc{DSDEPROJ} thermal profiles of CL1821+643 of \citet[][]{Russell10} in Sec. \ref{sec:s2.3_Russell10_descriptionAndTreatment}. These were inferred by using the \texttt{scipy.optimize.curve\_fit} algorithm for non-linear least squares minimisation\footnote{See \url{https://docs.scipy.org/doc/scipy/reference/generated/scipy.optimize.curve_fit.html}}. In Sec. \ref{sec:s2.3_Russell10_descriptionAndTreatment}, all best-fit free parameter values have been quoted with their corresponding standard deviation error. We note that performing the fits to the thermal profiles on the basis of $\chi^{2}$ minimisation only affects the survival probability curves at the $\lesssim \ 0.5\%$ level and hence does \textit{not} change our inferred constraints on ALPs. We note that our motivation behind Eq. \ref{eq:h1821_PRESSUREprofileFits}, defining our predicted pressure profile within CL1821+643 (orange curve in the upper panel of Fig. \ref{figure:VerticallyStackedFig_H1821_profiles}), was to find a model-independent description of $P_\mathrm{th}(r)$, consisting of the lowest possible number of free parameters. 

Our approach is consistent (yet model-independent) with the predictions inferred by the universal cluster pressure profile of \citet[][]{MArnaud2010}, extracted from fitting the \textsc{REXCESS} sample of 33 nearby ($z < 0.2$) clusters, of masses $\mathrm{log}(M_{500}/M_\odot) = 14 - 15$. However, they note that depending on their mass and thermodynamic properties, several clusters can depart from the inferred self-similar relations, all of which scale with cluster mass. Given the implicit independence of our fit to $P_\mathrm{th}(r)$ on $M_{500}$ (Eq. \ref{eq:h1821_PRESSUREprofileFits}), we do not expect this to have propagated into our analysis.

Although not needed directly in our analysis, we can use our best fits to the thermal properties of the host cluster to make an estimate of the overdensity mass $M_{500}$. We find the pressure gradient exerted on a gas cell $\mathrm{d}P(r)/\mathrm{d}r$ by finding the differential of Eq. \ref{eq:h1821_PRESSUREprofileFits} with respect to $r$, which denotes the co-moving distance to the cluster centre. Under the assumption of hydrostatic equilibrium, the latter is equivalent to the gravitational force exerted by the enclosed mass within the cluster up to $r$, $M(<r)$, onto such gas cell. Assuming a fully ionised plasma, to first order, the gas density is given by $\rho_\mathrm{g}(r) = m_\mathrm{p} \times n_\mathrm{e}(r)$, where $m_\mathrm{p}$ is the proton rest mass and $n_\mathrm{e}(r)$ is given by our best fit to the ICM electron number density of the host cluster (see Sec. \ref{sec:s2.3_Russell10_descriptionAndTreatment}). In line with the definition of $M_{500}$ provided in Sec. \ref{sec:s2.3_Russell10_descriptionAndTreatment}, and assuming a set of cosmological parameters consistent with those assumed throughout our spectral quasar analysis, we predict a cluster mass of $M_{500} \sim 3.9 \times 10^{14} M_\odot$. This estimate is roughly consistent with the cluster mass value previously suggested by a catalogue of thermal Sunyaev-Zel'dovich clusters \citep[][]{Planck13_tSZCatalogue}, $M_{500} = 6.3 \times 10^{14} \ M_\odot$.

\section{Adaptive treatment for resonances}
\label{sec:appB_adaptiveTreatmentForResonances}

In Sec. \ref{sec:s3_ALPSurvivalCurves}, the method by which we have generated a grid of photon-ALP mixing curves across ALP parameter space is described. In an early stage of this project, we noted that several of the high mass photon-ALP mixing curves we had produced (at $m_\mathrm{a} \lesssim 10^{-10.8} \ \mathrm{eV}$) presented significantly smooth or high-frequency oscillatory features at low energies. In some cases, we unexpectedly found these to be quite successful at describing the underlying structure of the residuals of our baseline quasar model (lower panel of Fig. \ref{figure:myModFiducial}). Given that we approximate the number density (and therefore $\omega_\mathrm{pl}$) to be constant across each cell, if the close-to-resonant photon-ALP inter-conversion regime ($\omega_\mathrm{pl}(r) \sim m_\mathrm{a}$) is met within a specific cell, this can translate into an artificial enhancement of the total photon-ALP mixing probability. In reality, quanta cross resonances near-instantaneously, although in practice, our early numerical treatment falsely produced prolonged close-to-resonant path lengths. We note that in gaussian random field (GRF) models describing the turbulent magnetic field, this artificial prolongation of the resonant term is less significant since no cell length is pre-defined. GRF models have a finer resolution to capture small-scale field behaviour, compared with cell-based approaches. We refer to \citet[][]{meyer2021} for a discussion on the latter, based on the divergence-free, homogeneous and isotropic GRF model presented in \citet[][]{Meyer2014_GRF}.
 
For this reason, we have implemented an adaptive treatment to deal with regions where the plasma frequency comes close to a resonance with the ALP mass, $m_\mathrm{a}$. For each \textit{Seed} and value of $m_\mathrm{a}$, we identify if the condition $\log(\omega_\mathrm{pl})-\log(m_\mathrm{a}) < \delta_\mathrm{res}$ is met in a cell, where $\delta_\mathrm{res}$ is a threshold value. A value of $\delta_\mathrm{res}=0.1$ was found to give reasonable and conservative results. We then split the identified cell into ten subsequent sub-cells and re-compute $n_\mathrm{e}(r)$ at the central value of each sub-cell, following Sec. \ref{sec:s3_ALPSurvivalCurves}, but keep the orientation and strength of the magnetic field fixed to the original value in the parent cell. Furthermore, within each problematic cell, we excise the sub-cell spanning the resonance condition ($\omega_\mathrm{pl} = m_\mathrm{a}$) from the calculation of $P_\mathrm{a\gamma}$ within the original (now fragmented) cell. We find this adaptive treatment to be effective in removing any artificial `boosting' of the ALP signal in our ALP-photon mixing models featuring $m_\mathrm{a} \sim \omega_\mathrm{pl}$. Although fairly conservative, our treatment could be improved by choosing the value of $\delta_\mathrm{res}$ in a more rigorous way, or by including the actual resonance calculation; we refer to \citet[][]{Marsh2021_Fourier} for a detailed discussion of the resonant regime. \citet[][]{Marsh2021_Fourier} also highlight the importance of the non-resonant terms in the calculation of photon-ALP mixing as they generically contribute equally, or to a higher extent than resonance.

\section{Minimum energy argument and equipartition field}
\label{sec:app_c_equipartition_derivation}
We revisit the synchotron minimum energy argument following \citet[][]{RybickiAndLightman}, who employ the \textit{cgs} unit convention.

As explained in Sec. 3.2 of \citet[][]{GOVONIandFERRETTI_2004}, the minimisation of the total energy density $U_\mathrm{tot}$ of a population of synchrotron-emitting electrons is achieved when the thermal and magnetic energy densities ($U_\mathrm{e}$ and $U_B$, respectively) are related by $U_\mathrm{e}$/$U_B = 4/3(1+k_\mathrm{prot})^{-1}$. Mathematically, this follows as: 
\begin{equation}
    \centering 
    \label{eq:Utot_from_equipartition}
U_\mathrm{tot} = (1 + k_\mathrm{prot} + 4/3) U_B \equiv (1 + k_\mathrm{prot} + 4/3) {\Tilde{B}}^{2} / {8\pi},
\end{equation}where $k_\mathrm{prot}$ is the ratio of the energy density stored in relativistic protons to electrons. Therefore, $\widetilde{B}$ in Eq. \ref{eq:Utot_from_equipartition} can be regarded as the field intensity required by the minimum energy argument. We note that we have ignored the contribution of inverse Compton emission to $U_\mathrm{tot}$.

On the other hand, we refer to Eq. 6.36 in \citet[][]{RybickiAndLightman}, which returns the total power per unit volume per unit frequency of synchrotron emission, $P_\mathrm{tot}(\omega$), which is a function of the frequency of observation $\nu$ of the observed synchrotron emission, since $\omega = 2\pi \nu$. Importantly, $P_\mathrm{tot}(\omega)$ depends on the normalisation constant of the synchrotron spectrum of the population of relativistic electrons, defined by Eq. 6.20b of \citet[][]{RybickiAndLightman}, i.e. $\mathrm{d}N(\gamma)/\mathrm{d}\gamma = C {\gamma}^{-p}$. The latter expression determines the number density of relativistic electrons with Lorentz factors in the range $\gamma \rightarrow \gamma + \mathrm{d}\gamma$, where $\alpha(p) \equiv \alpha = (p - 1)/2$ is the spectral index of the observed synchrotron spectrum. Using Eq. \ref{eq:Utot_from_equipartition}, we eliminate the dependence of $P_\mathrm{tot}(\omega)$ on $C$ to find an expression for $P_\mathrm{tot}(\omega, \widetilde{B})$.

For $p > 3$ (refer to Sec. \ref{subsubsection:MinEnergyArgument}), following on from the arguments above, we find the field intensity required by the minimum energy argument, $\widetilde{B}$, to be: 
\begin{equation}
    \centering 
    \label{eq:Beq_fromRybickiandLightman}
    \Tilde{B} = G { \biggl[\frac{8 \pi p}{{\gamma_\mathrm{min}}^{p}} \times \frac{P_\mathrm{tot} (\omega)/f \times {m_\mathrm{e} c}^{2} \times {[I(\omega)]}^{\alpha} }{\Gamma_\mathrm{1,2}(p) \times 4/3 \times H(\alpha)} \biggl]}^{-(1+3\alpha)},
\end{equation} where $m_\mathrm{e}c^{2}$ is the electron rest-mass energy, $\gamma_\mathrm{min}$ is the minimum Lorentz factor that dominates the low-energy cutoff of the synchrotron spectrum, and $f$ is the effective filling factor. Moreover, $\Gamma_{1,2}(p)$ is the product of the two gamma functions appearing in Eq. 6.36 in \citet[][]{RybickiAndLightman}, $H(\alpha)$ is its prefactor, which depends on the synchrotron spectral index, and $I(\omega)$, its $\omega$-dependent term (where we account for the K-correction). The constant $G = A(\xi)(1 + k_\mathrm{prot})^{2/7}$ depends on both $k_\mathrm{prot}$ \citep[see Eq. 25 of][]{GOVONIandFERRETTI_2004} and on a trigonometric function $A(\xi)$ of the pitch angle of synchrotron emission, $\xi$.

We note that throughout the calculations we refer to in Sec. \ref{subsubsection:MinEnergyArgument}, for simplicity, we take all trigonometric factors that depend on the pitch angle of synchrotron emission in Eq. \ref{eq:Beq_fromRybickiandLightman} to be $\mathrm{sin(\xi)} \rightarrow 1/2$. Finally, we convert the observed flux density of the diffuse synchtrotron emission quoted in \citet[][]{CL1821+643_Bonafede14} at $\nu_\mathrm{obs} = 323 \ \mathrm{MHz}$, $F_{\nu_\mathrm{obs}}$, into an energy density via $P(\omega_\mathrm{obs}) = F_{\nu_\mathrm{obs}} \times 4 \pi [d_\mathrm{Lum}(z = 0.299)]^{2} / V$, where $d_\mathrm{Lum}$ is the luminosity distance of the quasar and $V$ is the volume of the radio halo, which we assume to be that of an ellipsoid of diameters $890 \ \mathrm{kpc} \times 450 \ \mathrm{kpc} \times 670 \ \mathrm{kpc}$. We additionally use this argument on the observed flux density of the central FR-I radio structure at $5 \ \mathrm{GHz}$ \citep[following][]{BlundellandRawlings_2001_FRI_inCL1821+643}, assuming that $V$ corresponds to the volume of a sphere of radius $140 \ \mathrm{kpc}$ \citep[see Fig. 2 of][]{BlundellandRawlings_2001_FRI_inCL1821+643}.

\label{lastpage}
\end{document}